\documentclass[twocolumn]{aastex62}
\usepackage{natbib}
\usepackage{longtable}
\usepackage{multirow}
\usepackage{amsmath}
\usepackage{makeidx}
\usepackage{xcolor}

\submitjournal{ApJ}

\shorttitle{}
\shortauthors{}
\begin{document}

\title{EXPLAINING THE MULTIPLE POPULATIONS IN GLOBULAR CLUSTERS BY MULTIPLE EPISODES OF STAR FORMATION AND ENRICHMENT WITHOUT GAS EXPULSION FROM MASSIVE STAR FEEDBACK}

\correspondingauthor{Young-Wook Lee}
\email{ywlee2@yonsei.ac.kr}

\author{Jenny J. Kim}
\affil{Center for Galaxy Evolution Research \& Department of Astronomy, Yonsei University, Seoul 03722, Korea}

\author{Young-Wook Lee}
\affil{Center for Galaxy Evolution Research  \& Department of Astronomy, Yonsei University, Seoul 03722, Korea}

\begin{abstract}

In order to investigate the origin of multiple stellar populations found in globular clusters (GCs) in the halo and bulge of the Milky Way, we have constructed chemical evolution models for their putative low-mass progenitors. In light of recent theoretical developments, we assume that supernova blast waves undergo blowout without expelling the pre-enriched ambient gas, while relatively slow winds of massive stars, together with the winds and ejecta from low to high mass asymptotic-giant-branch stars, are all locally retained in these less massive systems. Interestingly, we find that the observed Na-O anti-correlations in metal-poor GCs can be reproduced when multiple episodes of starburst and enrichment are allowed to continue in these subsystems. A specific form of star formation history with decreasing time intervals between the successive stellar generations, however, is required to obtain this result, which is in good agreement with the parameters obtained from synthetic horizontal-branch models. The ``{mass budget problem}" is also much alleviated by our models without ad-hoc assumptions on star formation efficiency, initial mass function, and the preferential loss of first-generation stars. We also applied these models to investigate the origin of super-He-rich red clump stars in the metal-rich bulge suggested by \citet{lee15}. We find that chemical enrichment by the winds of massive stars can naturally reproduce the required strong He enhancement in metal-rich subsystems. Our results further underscore that gas expulsion or retention is a key factor in understanding the multiple populations in GCs.

\end{abstract}

\keywords{Galaxy: formation ---
  globular clusters: general ---
  stars: abundances
  }

\section{Introduction} \label{sec:intro}
Ever since the first discovery of the multiple and discrete sequences among red-giant-branch (RGB) stars in $\omega$ Cen \citep{lee99}, most of the globular clusters (GCs) are now known to host multiple stellar populations with different He and light element abundances \citep[][and references therein]{gratton12, piotto15, lim15, milone17}. Spectroscopic studies revealed that almost all of the GCs investigated hitherto show an anti-correlation between Na and O abundances with little or no dispersion in heavy element abundances such as Fe and Ca \citep[see, e.g.,][]{carretta09b, carretta15}. Such anti-correlated abundance trends are also observed between N and C \citep[e.g.,][]{cohen05} as well as Al and Mg \citep[e.g.,][]{meszaros15}. These abundance patterns and the discrete sequences of stars in the Hertzsprung-Russell (HR) diagram are found not only in RGB stars but also in main-sequence stars \citep[e.g.,][]{gratton01, cohen02, briley04, bedin04}, suggesting primordial origin for this phenomenon.

Recently, the observed double red clump (RC) and the two populations of RR Lyrae stars in the Milky Way (MW) bulge are suggested to be another manifestation of the same He-enhanced multiple population phenomenon as is observed in GCs \citep{lee15, lee16, joo17}. If confirmed, this would indicate that proto-GCs played a key role in the bulge formation by providing He-enhanced stars (as well as He-normal stars) to the bulge field in the early phase of the MW formation. The main underlying hypothesis is that He abundance of the second-generation stars (G2) increases strongly with metallicity following a steep He enhancement parameter ($\Delta Y/\Delta Z \approx 6$), while the first-generation stars (G1) obey the standard enrichment ratio ($\Delta Y/\Delta Z \approx 2$). Direct evidence for this multiple population origin of the double RC phenomenon has been reported by \citet{lee18}, where a statistically significant difference in CN-band strength has been detected between the stars in the two RC regimes.

The self-enrichment scenario is commonly invoked to explain the multiple population phenomenon in GCs. This scenario assumes that stars enhanced in He, N, and Na (depleted in C and O) form from the gas that has been polluted by processed materials ejected by G1. Stars experiencing proton-capture and CNO-cycle at high temperature have been suggested as the source of such abundance patterns. Popular candidates for these polluting stars are massive asymptotic-giant-branch stars \citep[massive AGB;][]{dantona04, dercole10, dantona16}, fast-rotating-massive stars \citep[FRMS;][]{decressin07a, decressin07b, krause13}, massive-interacting-binaries \citep[MIB;][]{demink09, bastian13}, and supermassive stars \citep[SMS;][]{denissenkov14, gieles18}.

  None of these models are, however, successful in explaining all the observed properties of GCs \citep{renzini15, bastian17}. For example, at least half of the stars in GCs appear to be enriched in He, N and Na \citep[see, e.g.,][]{carretta09b, bastianlardo15, milone17}, while the ejecta from proposed candidates constitutes only a small fraction of the total initial mass of G1 when weighted with a canonical stellar initial mass function (IMF). The predominance of G2, therefore, cannot be explained without additional ad-hoc assumptions. In order to overcome this ``{mass budget problem}" \citep{renzini08}, most of the suggested models appeal either to the extremely top-heavy IMF for G1 or to the preferential removal of G1 from GCs which are assumed to be $10-100$ times more massive at the time of formation \citep{dantona04, prantzos06, dercole08, decressin07a, conroy12}. The latter, however, is not supported by some observations of metal-poor GCs and halo stars in dwarf galaxies, which suggest proto-GCs are not likely to be more than $4-5$ times  massive compared to the present mass \citep{larsen12, larsen14}. Moreover, in order to reproduce the full-range of the anti-correlated abundance patterns, Na-O anti-correlation in particular, dilution with pristine gas (which has the same chemical composition as G1) is essential in most of the suggested models. However, it is considered to be questionable whether such a process can naturally explain the discrete distribution of subpopulations observed in photometric and spectroscopic studies \citep{bastian15}. The readers are referred to \citet{renzini15} and \cite{bastian17} for recent reviews on this topic.
  
The models invoking AGB as the source of the processed gas assume that the energy injection by the winds of massive stars (WMS) and type II supernovae (SNe) can form a powerful cluster wind which would effectively remove the remaining natal gas together with the SN ejecta \citep{calura15, dercole16}. The formation of G2 then takes place using the gas that has been entirely enriched by AGB which has no spreads in heavy element abundances. However, recent theoretical studies, with more realistic treatment on the central gas density and density gradient in a proto-GC, suggest that SN ejecta can preferentially escape the system without expelling the remaining gas, while the WMS are mostly retained (\citealt{tenorio-tagle15, silich17, silich18}; see also \citealt{krause12, krause16}). Some recent observations in infrared, sub-millimeter, and radio across massive star clusters have also challenged the notion that the feedback from massive stars would drive a cluster wind \citep[see, e.g.,][]{turner17, oey17, cohen18}. In this paper, we follow the chemical evolution in a proto-GC with the assumption that SN ejecta can preferentially escape the system without affecting the remaining gas\footnote{A similar treatment on the fate of SN ejecta has also been adopted by \citet{romano10} in their investigation of the massive GC $\omega$ Cen, and by some models invoking extra mass-loss from massive stars as they require  formation of G2 to proceed without being contaminated or interfered by subsequent SNe explosions \citep{bastian13, krause13}.}. The purpose of this paper is to show that the observed Na-O anti-correlation in GCs and the presence of super-He-rich stars in the MW bulge can be naturally reproduced when we adopt such an assumption on the fate of SN ejecta together with multiple star forming episodes and continuous enrichments by successive generations.

\section{Chemical evolution models} \label{sec:cemodel}
\subsection{Basic assumptions} \label{subsec:assumption}

\begin{figure}
\includegraphics[scale=0.75]{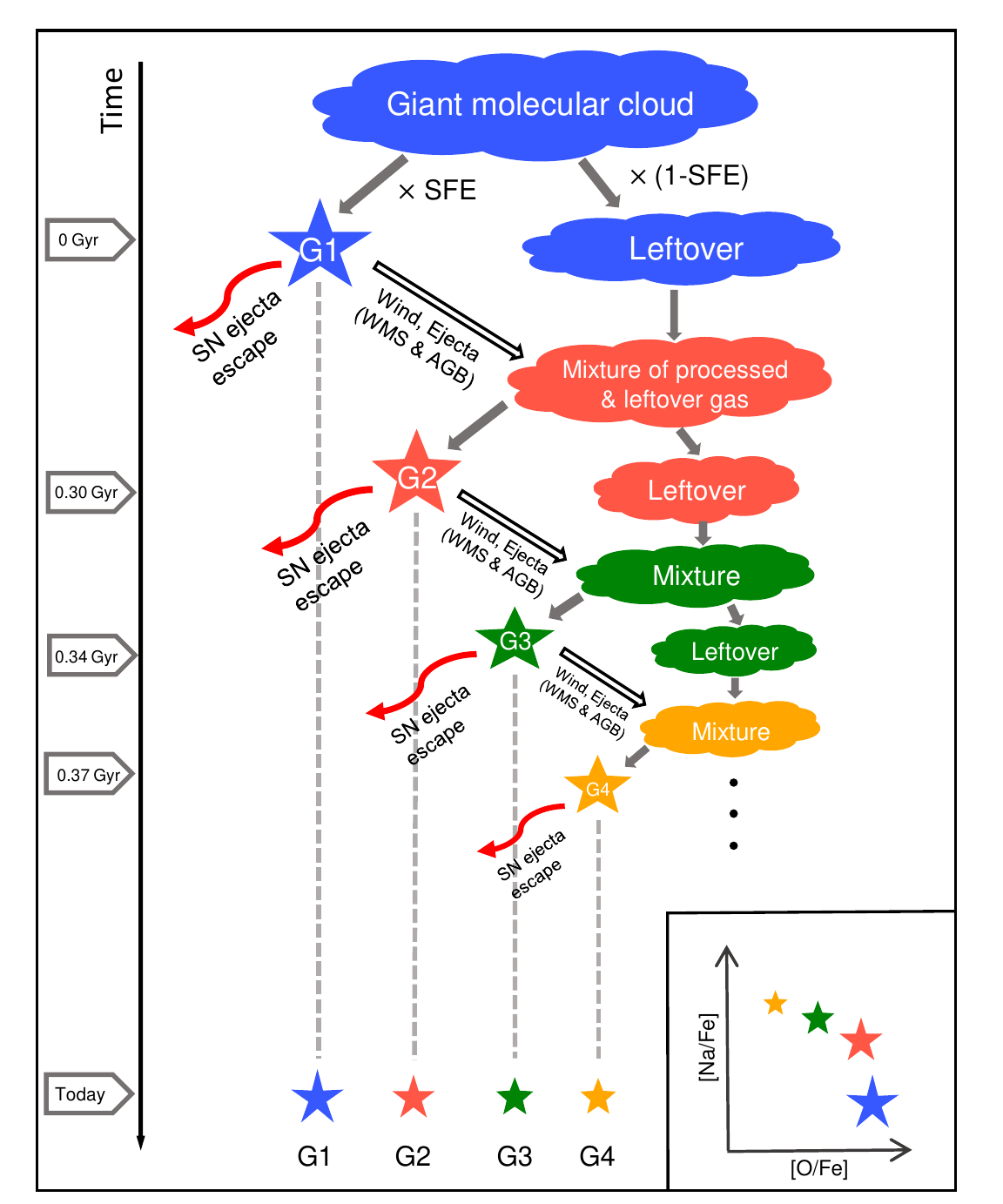}
\centering
\caption{Schematic illustration of our model for the formation of multiple populations in a proto-GC. Chemical enrichment process is shown with an example of star formation history corresponding to our best-fit model of M4 (see Table~\ref{tab:m4m5}). The colors reflect the chemical composition of gas and stars: blue, red, green, and orange correspond, respectively, to G1, G2, G3, and G4, and their placements on the Na-O plane are also shown.
\label{schematic}
}
\end{figure}

Our models differ from previous approaches in that we assume (1) SNe blast waves undergo blowout without expelling the pre-enriched ambient gas in the proto-GCs; and (2) discrete star forming episodes beyond G2, to the third, fourth, and later generations (G3, G4 \dots) with continuous enrichments by successive generations. These processes would continue until no enough gas is left in the system after the successive star formations. The explosions of Type Ia SNe \citep{dercole08, dantona16} and the ram pressure exerted in the orbital motion \citep{conroy11} would also help to clear out the remaining gas from the system. The chemical evolution in our model is, therefore, dictated by the ejecta from AGB and WMS. Instead of relying on a specific mechanism (such as fast rotation or interacting binaries) or mass range of stars, we adopt standard stellar evolutionary models for the WMS and AGB with fully populated IMF commonly invoked in galactic chemical evolution models \citep[see, e.g.,][]{mcwilliam08, cote13}. Figure~\ref{schematic} shows a schematic representation of our models with the basic assumptions described above. Since multiple episodes of star formation are allowed in our models, the gas in a proto-GC becomes more and more enriched by the processed materials from WMS and AGB ejecta of successive generations, making it possible to reproduce the most extreme abundances in He and Na at the latest generation. This is fundamentally different from the approaches taken by most of the previous models where only G1 is considered as a source for the chemical enrichment\footnote{A successive enrichment process has also been suggested by \citet{elmegreen17} in his investigation based on interacting massive stars and MIB. The timescale for the formation of multiple populations in his models, however, is  two orders of magnitude smaller than those predicted in our models because SNe explosions are assumed to clear out the remaining gas preventing further star formation.}. In such models, stars with the most extreme abundances are assumed to form out of the gas that is almost completely made up of the processed materials ejected by G1.

\subsection{Model construction} \label{subsec:construction}
Our computation is based on the basic formalisms of chemical evolution pioneered by \citet{tinsley80}. Starting from a giant molecular cloud of mass $M_{GMC}$ with the present-day metallicity of a specific GC being considered, G1 are assumed to form instantaneously at time $t_{G1}$ with a star formation efficiency (SFE) of 0.6. This quantity is consistent with the predicted value from theoretical studies of GC formation in dense clouds \citep{elmegreen97}, and is similar to  the commonly adopted values (0.4--0.5) in the models for the multiple population phenomenon of GCs \citep[e.g.,][]{conroy12, calura15}. Then, the initial mass of the G1 and mass of the leftover gas after the first episode of star formation are expressed as:
\begin{equation}
M_{G1} = M_{GMC}  \times SFE,
\end{equation}
\begin{equation}
M_{leftover}(t_{G1}) = M_{GMC}  \times (1-SFE).
\end{equation}

Subsequent star forming episodes beyond G1 take place using the gas that is a mixture of the leftover gas and the processed materials ejected from previous generations which are assumed to mix instantaneously. The initial stellar mass of the n'th-generation (G($n$)) and the mass of the leftover gas at the formation time, $t_{G(n)}$, are then expressed as 
\begin{equation}
M_{G(n)} = [M_{leftover}(t_{G(n-1)}) + E(t_{G(n)})]  \times SFE,
\end{equation}
\begin{equation}
\begin{aligned}
M_{leftover}(t_{G(n)}) = {} &\ [M_{leftover}(t_{G(n-1)}) + E(t_{G(n)})] \\
& \times (1-SFE),
\end{aligned}
\end{equation}
where $M_{leftover}(t_{G(n-1)})$ is the mass of the leftover gas after the formation of G($n-1$), and $E(t_{G(n)})$ is the total mass of the processed materials ejected by previous generations from time $t_{G(n-1)}$  to $t_{G(n)}$. 

For all stellar generations, we assume a fixed SFE and a single-slope IMF, $\phi(m)$, with masses ranging from 0.1 to 120$M_{\sun}$. We allow the IMF slope, $s$, to vary from 1.8 to 2.1 depending on the GC being considered. This is slightly top-heavier compared to the value of \citet[][s = 2.35]{salpeter55}, however, is fully consistent with recent observations at masses more than 1$M_{\sun}$ \citep[see the compilation by][]{cote16}. The $E(t_{G(n)})$ is then calculated by summing up the contributions from generations that have formed before $t_{G(n)}$. The mass of the ejected materials by a single stellar generation from time $t$ to $t+\delta t$ is expressed as
\begin{equation}
\int_{M_{t+ \delta t}}^{M_{t}} m_{eje}(m) \phi(m)dm,
\end{equation}
where $M_{t}$ and $M_{t+ \delta t}$ indicate the initial masses of a star with the lifetimes of $t$ and $t+\delta t$, respectively, and $m_{eje}(m)$ is the mass of WMS or AGB ejecta depending on the initial mass $m$. The stellar lifetimes are adopted from \citet{portinari98} with corrections for the effects of He enhancement \citep[from][]{karakas14} for the G2 and later generation stars.

As for the variation of the elemental abundances of the gas in a proto-GC, the total mass of the element $i$ ejected by a single stellar generation between time $t$ and $t+\delta t$ is expressed as 
\begin{equation}
\int_{M_{t+ \delta t}}^{M_{t}} [X_{i}m_{eje}(m) + mp_{i}(m)] \phi(m)dm,
\label{eq6}
\end{equation}
where $X_{i}$ denotes the abundance of element \textit{i} of the generation being considered, and $p_{i}(m)$ is the stellar yield which is defined as a mass fraction of a star of mass $m$ that is newly converted to element \textit{i} and then ejected. For the extreme subpopulations where C and O are already extremely depleted, the ejected masses of C and O could become unrealistically negative and cause a numerical error for stars within a certain mass range ($\sim$$6-8M_{\sun}$ for AGB ejecta, and $\sim$$30-50M_{\sun}$ for WMS)\footnote{Obviously, the ejected mass of an element cannot be negative as this would indicate that stars burn more amount of the element than what was initially present.}. In these cases, we apply corrections to the ejected masses so that they are always maintained to be greater than or equal to zero. Consequently, the predicted abundances of C and O for G5 and later generations would be more uncertain. By summing the equation~(\ref{eq6}) for all individual generations with assumed ages, initial masses, and chemical compositions, $E_{i}(t_{G(n)})$, the total ejected mass of element $i$ by previous generations from time $t_{G(n-1)}$  to $t_{G(n)}$, can be obtained. The chemical composition of G1 is adopted from \citet{grevesse93} with changes in the alpha elements (including O) and Na to match the observed placement of G1 on the Na-O plane, and the He content of G1 is obtained by following the standard He to metal enrichment ratio ($\Delta Y/\Delta Z = 2$). The adopted stellar yields are described in the following subsection. As the formation of $G(n)$ takes place using the gas that is a mixture of the ejecta and the remaining gas, the abundance of element $i$ of $G(n)$ can be computed using the expression:
\begin{equation}
X_{G(n),i} = \frac{E_{i}(t_{G(n)})+X_{G(n-1),i}M_{leftover}(t_{G(n-1)})}{E(t_{G(n)})+M_{leftover}(t_{G(n-1)})},
\end{equation}
where $X_{G(n-1),i}$ denotes the abundance of element \textit{i} of $G(n-1)$. When calculating population ratios of G1, G2 and later generations, we considered stars in the mass range from the main-sequence turn-off to the tip of RGB determined from the Yonsei-Yale ($Y^{2}$) isochrones with different values of He content \citep{yi08} as spectroscopic observations are usually made for RGB stars.
  
\begin{deluxetable*}{ccccc}
\setlength{\tabcolsep}{0.15in}
\tablecaption{Fraction of the total ejected mass of each element by WMS and AGB (with different mass ranges) as a function of metallicity \label{tab:contribution}}
\tablehead{
\colhead{Element} & \colhead{f(WMS: $120-9M_{\sun}$)} & \colhead{f(AGB: $9-5M_{\sun}$)} & \colhead{f(AGB: $5-3M_{\sun}$)} & \colhead{f(AGB: $3-2M_{\sun}$)}
}
\startdata
[Fe/H] = $-$0.1 & & & &\\
He & \textbf{0.69} & 0.12 & 0.10 & 0.08\\
C & \textbf{0.92} & 0.00 & 0.02 & 0.06 \\
N & \textbf{0.66} & 0.16 & 0.13 & 0.05 \\
O & \textbf{0.53} & 0.15 & 0.17 & 0.15 \\
Na& \textbf{0.47} & 0.26 & 0.21 & 0.05 \\
\noalign{\vskip 1.8mm}    
\hline
[Fe/H] = $-$0.9 & & & & \\
He& \textbf{0.44} & 0.23 & 0.19 & 0.14 \\
C & 0.03 & 0.00 & 0.10 & \textbf{0.86} \\
N & 0.35 & 0.23 & \textbf{0.37} & 0.05 \\
O & 0.21 &0.10 & 0.19 & \textbf{0.49}\\
Na& 0.33 & 0.17 &\textbf{0.37} & 0.13\\
\noalign{\vskip 1.8mm}  
\hline
[Fe/H] = $-$2.0 & & & & \\
He& 0.29 & \textbf{0.30} & 0.25 & 0.16\\
C & 0.00 & 0.01 & 0.04 & \textbf{0.95}\\
N & 0.08 & 0.12 &\textbf{0.50} & 0.30 \\
O & 0.06 & 0.03 & 0.19 & \textbf{0.73} \\
Na& 0.09 & 0.09 & \textbf{0.45} & 0.37\\
\noalign{\vskip 1.8mm}  
\enddata
\tablecomments{Stellar evolutionary phases mainly responsible for the enrichment of the considered elements are highlighted in boldface. The IMF slope of 2.1 is assumed. }
\end{deluxetable*}

\subsection{Stellar yields} \label{subsec:yields}
\subsubsection{Winds of massive stars} \label{subsubsec:wms}
Before massive stars explode as SNe, they enrich the interstellar medium by means of radiatively-driven stellar winds. In the metal-poor regime, WMS are mainly composed of H-burning products which are enhanced in He and N but depleted in C and O. However, in the metal-rich regime, He-burning products, C and O, are also provided by WMS because of the extremely high mass-loss rate experienced by metal-rich massive stars in the Wolf-Rayet stage \citep{maeder92, portinari98, meynet08}. Such a stronger wind also provides larger amounts of newly formed He and N as well (see Table~\ref{tab:contribution}). 

We adopt metallicity dependent yields of WMS from \citet{portinari98} which covers the wide ranges of parameters required in the calculations such as metallicity, stellar mass, and chemical yields of WMS for various elements. Since Na yields are not available for the WMS from this study, we made use of the total ejected mass of WMS in the form of Na provided by chemical evolution models of \citet{cote13}. Also, the treatment of stellar rotation is not included in the models of \citet{portinari98} which is known to play a key role in the evolution of massive stars \citep{meynet17}. In Section~\ref{sec:lc18} below, we have therefore explored the effects of rotation by adopting the latest calculation of massive star evolution by \citet{limongi18}.

In our models, G2 and later generations are predicted to be enhanced in He and have variations in light element abundances with an overall increase, in most cases, in CNO content (see Table~\ref{tab:othergcs} below). Unfortunately, none of the models are available in the literature for the yields from stars with such enhancements, and therefore, the effects of these enhancements are not reflected in our models as in the previous models \citep[see, e.g.,][]{choi07, romano10}. However, the effects of CNO and He enhancements are likely to be minimal because the increased mass-loss rate from CNO enhancement \citep{vink01} would mostly be canceled out by the decreased lifetime from He enhancement.\\

\begin{deluxetable*}{cl}
\setlength{\tabcolsep}{0.2in}
\tablecaption{Main sources of elements in our models\label{tab:source}}
\tablehead{
\colhead{Element} & \colhead{Main Source}
}
\startdata
\noalign{\vskip 1.mm}
He & Enriched by massive AGB ${(5M_{\sun} {\lesssim} M_{ini} {\lesssim} 8M_{\sun})}$ and WMS \\
\noalign{\vskip 1.8mm}   
N, Na &  Enriched by intermediate mass AGB ${(3M_{\sun} {\lesssim} M_{ini} {\lesssim} 5M_{\sun})}$ and metal-rich WMS \\
\noalign{\vskip 1.8mm}   
C, O & Enriched by low mass AGB ${(M_{ini} {\lesssim} 3M_{\sun})}$ and metal-rich WMS\\
& Depleted by massive AGB ${(5M_{\sun} {\lesssim} M_{ini} {\lesssim} 8M_{\sun})}$ and metal-poor WMS \\
\noalign{\vskip 1.8mm}  
\enddata
\end{deluxetable*}

\subsubsection{Winds and ejecta from AGB} \label{subsubsec:agb}
The chemical composition of the AGB ejecta depends on the initial mass of a star. According to \citet{ventura13}, this is because the higher the mass of a star, the higher the temperature at the bottom of the convective envelope which would result in a stronger hot bottom burning. For this reason, the ejected materials from massive AGB with initial masses in the range of ${5M_{\sun} {\lesssim} M_{ini} {\lesssim} 8M_{\sun}}$ are mostly depleted in C and O along with substantial enhancement in He. The Na, however, is mainly enriched by intermediate mass AGB (${3M_{\sun} {\lesssim} M_{ini} {\lesssim} 5M_{\sun}}$) as a result of efficient Ne22 burning. In these stars, N is also efficiently produced from C. Low mass AGB ($M_{ini} {\lesssim} 3M_{\sun}$) are dominated by the third dredge-up without the effects from hot bottom burning; therefore, their ejecta is mainly enriched in C and O with an overall increase in CNO content.

We adopt the yields of AGB from \citet{ventura13} for Z = 0.0003 and 0.001, \citet{ventura14} for Z = 0.004, and \citet{dicriscienzo16} for Z = 0.02. The effects of He enhancement on the AGB yields are not included in our calculations as the parameter space available in the literature is rather limited. However, it appears to have only a negligible impact on our results as the changes in the yields are not significant within the range of He enhancement in a typical GCs \citep{karakas14, shingles15}. Table~\ref{tab:contribution} shows the result for the metallicity dependence of chemical composition ejected by stellar models embedded in our calculation. Table~\ref{tab:source} briefly summarizes the main sources of elements in our chemical evolution models.

\vskip 15.mm 

\begin{figure*}
\includegraphics[scale=.7]{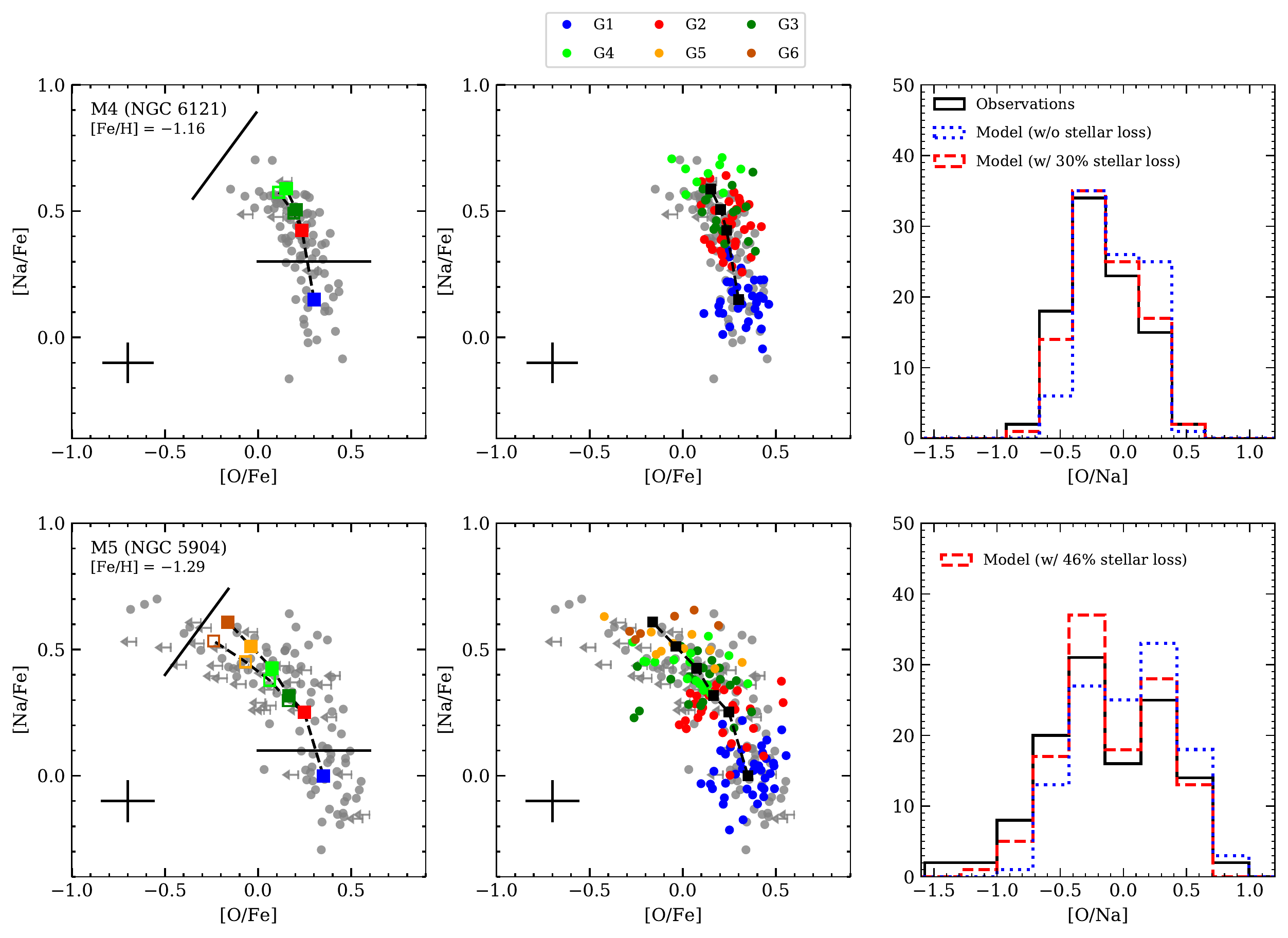}
\centering
\caption{Comparison of our chemical evolution models with the observed Na-O anti-correlations in M4 and M5 \citep[data in gray circles are from][]{carretta09a, carretta09b}. In the left panel, our model predictions are depicted by filled squares following the color scheme in the legend. The open squares are for the case where no contribution of G1 ejecta is assumed to the formation of G3 and later generations. For comparison, separations between ``{Primordial}" \& ``{Intermediate}" and ``{Intermediate}" \& ``{Extreme}" components defined by \citet{carretta09b} are shown as solid lines. The short arrows indicate upper limits on the observed [O/Fe] abundances. In the middle panel, our model prediction including observational errors is shown by circles following the color scheme in the legend by employing the population ratio after the stellar loss. The filled squares in the left panel are repeated here by black squares. In the right panel, predicted [O/Na] histogram from our model is compared with observations. To better match the observed histogram, some fraction of the stars made up of earlier generations were assumed to be preferentially lost, and the required fraction is indicated in the legend. Note that a specific shape of SFH is required to reproduce the observation (see Table~\ref{tab:m4m5}). \label{m4m5}}
\end{figure*}

\section{Origin of N\MakeLowercase a-O anti-correlation in globular clusters} \label{sec:NaO}

\subsection{Models for M4 and M5\label{subsec:M4M5}}

In spectroscopic studies, the multiple stellar populations of GCs are characterized by the Na-O anti-correlation. Here, we employ our models described above to reproduce the observed Na-O anti-correlations of the GCs M4 (NGC~6121) and M5 (NGC~5904), both of which are well established from relatively large samples of RGB stars \citep{carretta09a, carretta09b}. As is shown in Figure~\ref{m4m5}, the Na-O anti-correlation observed in M5 is rather extended with the presence of extremely O-poor stars, while that of M4 shows relatively modest extension, and could be considered as a typical GC. In our modeling, there are three input parameters, SFE, IMF slope, and star formation history (SFH; time intervals between the formations of successive generations). Among these, SFE and IMF slope can be inferred, to certain ranges, from observations and theoretical studies discussed above, while the SFH is considered to be a free parameter and allowed to vary without prior information. At given values of SFE (0.6) and IMF slope (s = $1.8-2.1$), the ``{best-fit}" model is obtained by varying the SFH until the model best matches the observed Na-O anti-correlation including the [O/Na] histogram. The effects of variations in these input parameters are discussed in the next subsection.

\begin{figure}
\includegraphics[scale=.7]{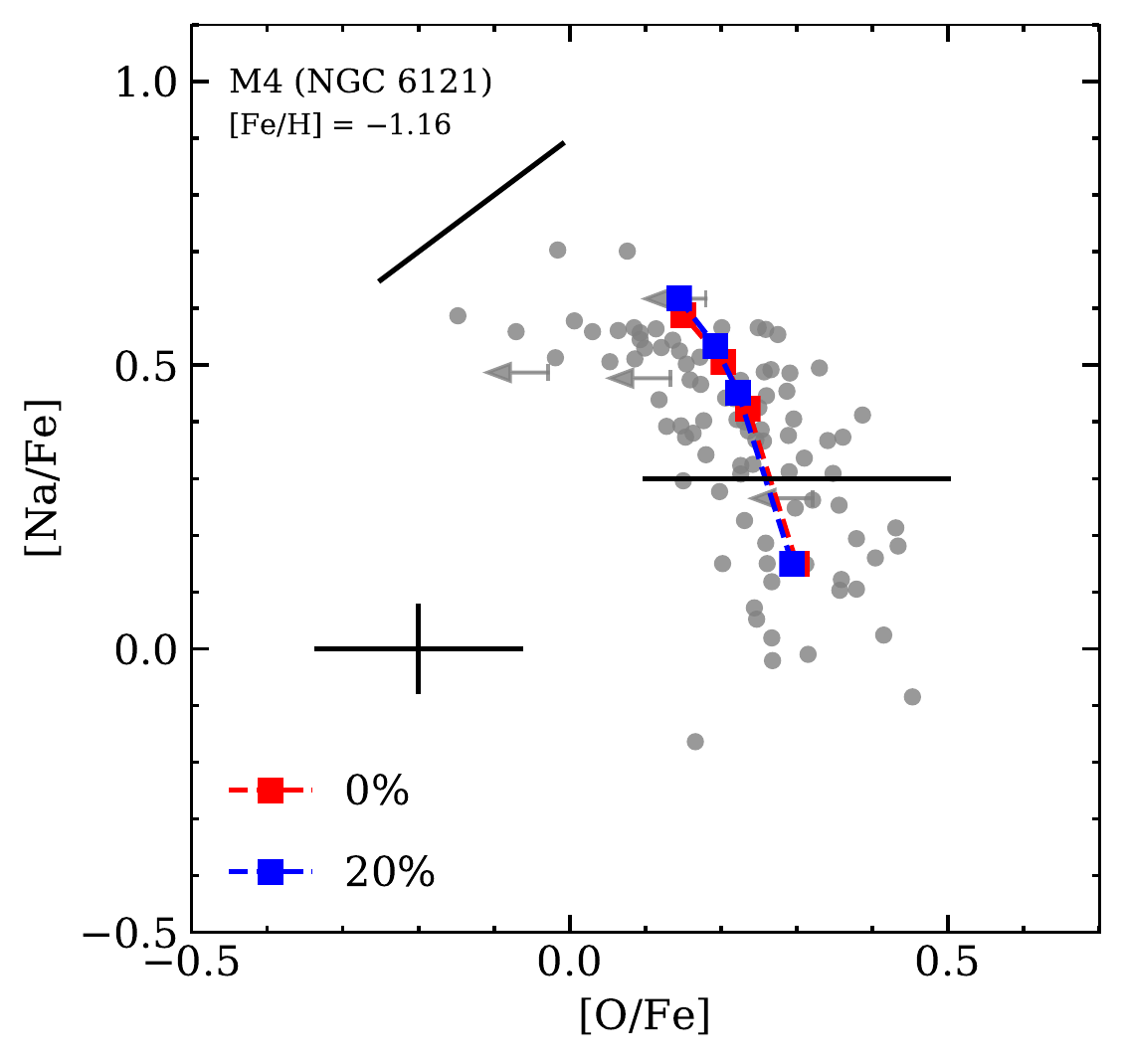}
\centering
\caption{Same as the left panel of Figure~\ref{m4m5} but the models are constructed with an assumption that  20\% of the ambient gas is entrained in the outflow driven by SN explosions (blue squares) which are compared with the models without this prescription (red squares). The predicted Na-O anti-correlation is little affected by this treatment.} \label{m4ent}
\end{figure}

The observed Na-O anti-correlations are compared with our best-fit models for M4 (with s = 2.1) and M5 (with s = 1.8) in Figure~\ref{m4m5}. For both GCs, a specific form of SFH with decreasing time intervals between successive generations is required to best match the observed anti-correlation (see Table~\ref{tab:m4m5}). This is because O is mostly depleted by massive AGB ($5M_{\sun} \lesssim M_{ini} \lesssim 8M_{\sun}$) and WMS, whereas Na is mainly enriched by relatively less massive AGB ($3M_{\sun} \lesssim M_{ini} \lesssim 5M_{\sun}$) as summarized in Table~\ref{tab:source}. Note that the Na-O anti-correlation of M5, which is relatively extended, requires a larger number of star forming episodes compared to that of M4. As for the population ratio, our models predict $\sim$50\% of the stars to be substantially enhanced in Na (see Table~\ref{tab:m4m5}) which is roughly consistent with the observed [O/Na] histograms as shown in the right panels of Figure~\ref{m4m5}. However, in order to better match these histograms, it is required to assume that $\sim$$30-50\%$ of the stars, preferentially made up of the earlier generations (G1 \& G2 for M4; G1, G2  \& G3 for M5), were lost, most likely due to the tidal shocks in the host galaxy disc \citep[e.g.,][]{kruijssen15}. Nevertheless, the required mass of the lost stars is an order of magnitude smaller than those required in previous models  \citep[see, e.g.,][]{dantona04, dercole08, decressin07a, conroy12}. Therefore, the mass budget problem is much alleviated by our models without ad-hoc assumptions on SFE, IMF, and the preferential loss of G1. Even if we assume that 20\% of the ambient gas is entrained in the outflow driven by SN explosions, the initial population ratio of G1 is increased only by 7\% and so that $\sim$40\% of the stars are still predicted to be substantially enhanced in Na without significant changes in Na and O abundances (see Figure~\ref{m4ent} and Table~\ref{tab:m4ent}).

If, as suggested by \citeauthor{dercole08} (\citeyear{dercole08}; see also \citealt{bekki11}), G2 and later generations form in more centrally concentrated environment compared to G1, the processed materials from G1 are not likely to fully contribute to the formation of later generations. In this regard, we also explored the case where none of the ejecta from G1 is retained when forming G3 and later generations (see the open squares in Figure~\ref{m4m5}). In this case, our models predict later generations to be more depleted in O (and C; see Figure~\ref{CN} below) as the CNO enhanced ejecta from low-mass AGB of G1 does not contribute to the chemical enrichment. 

\begin{deluxetable*}{ccccccccc}
\setlength{\tabcolsep}{0.1in}
\tablecaption{Results of our best-fit models for the GCs M4 and M5 \label{tab:m4m5}}
\tablehead{
\colhead{Population} & \colhead{Y} & \colhead{[Na/Fe]} & \colhead{[O/Fe]} & \colhead{[N/Fe]}  & \colhead{$\Delta$[CNO/Fe]} &\colhead{Fraction} & \colhead{Fraction} & \colhead{t(Gyr)} \\
\colhead{} &\colhead{}& \colhead{} & \colhead{} & \colhead{} & \colhead{} &\colhead{original} & \colhead{remaining} & \colhead{} 
}
\startdata
\multicolumn{3}{c}{\textbf{M4} ([Fe/H] = $-$1.16, s = 2.1)}& \\
\noalign{\vskip 1mm}  
G1 &  0.234 & 0.15 & 0.30 & 0.00 & 0.00  & 0.52 & 0.33 & 0.00 \\
G2 &  0.259 & 0.42 & 0.24 & 0.79 & 0.06  & 0.28 & 0.38 & 0.30 \\
G3 &  0.277 (0.277)& 0.51 (0.49)& 0.20 (0.19)& 0.94 (0.92)& 0.10 (0.07) & 0.13 & 0.19 & 0.34 \\
G4 &  0.299 (0.301)& 0.59 (0.57)& 0.15 (0.11)& 1.09 (1.05)& 0.15  (0.08) & 0.07& 0.10 & 0.37 \\
\noalign{\vskip 2mm}   
\hline
\multicolumn{3}{c}{\textbf{M5} ([Fe/H] = $-$1.29, s = 1.8)}&&\\
\noalign{\vskip 1mm}  
G1 &  0.233 & 0.00 & 0.35 &  0.0  &    0.0   & 0.45 & 0.33 & 0.0 \\
G2 &  0.272 & 0.25 & 0.25 &  0.90 &  0.07  & 0.26  & 0.23 & 0.20 \\
G3 &  0.305 (0.306) & 0.32 (0.30)& 0.17 (0.16) &  1.11 (1.10) &  0.12 (0.11) & 0.13 & 0.17 & 0.22 \\
G4 &  0.339 (0.341) & 0.43 (0.38)& 0.07 (0.06) &  1.27 (1.24) &  0.18 (0.16) & 0.08 & 0.14 & 0.24 \\
G5 &  0.373 (0.376) & 0.51 (0.45)&-0.04 (-0.07)&  1.38 (1.35) & 0.23 (0.20) & 0.05 & 0.08 & 0.25 \\
G6 &  0.401 (0.407) & 0.61 (0.53)&-0.16 (-0.24)&  1.47 (1.43) & 0.28 (0.23) & 0.03 & 0.05 & 0.26 \\
\noalign{\vskip 2mm}   
\enddata
\tablecomments{Quantity in parenthesis is the result of our calculation with no contribution of G1 ejecta to the formation of G3 and later generations. The relative difference of [CNO/Fe] is with respect to G1. The last column is the time elapsed since the formation of G1. }
\end{deluxetable*}

\begin{deluxetable*}{cccccccc}
\setlength{\tabcolsep}{0.15in}
\tablecaption{Result of our model for M4 constructed with an assumption that 20\% of the ambient gas is entrained in the outflow driven by SN explosions \label{tab:m4ent}}
\tablehead{
\colhead{Population} & \colhead{Y} & \colhead{[Na/Fe]} & \colhead{[O/Fe]} & \colhead{[N/Fe]}  & \colhead{$\Delta$[CNO/Fe]} &\colhead{Fraction}  & \colhead{t(Gyr)} \\
\colhead{} &\colhead{}& \colhead{} & \colhead{} & \colhead{} & \colhead{} &\colhead{original} &  \colhead{} 
}
\startdata
\multicolumn{3}{c}{\textbf{M4} ([Fe/H] = $-$1.16, s = 2.1)}& \\
\noalign{\vskip 1mm}  
G1 &  0.234 & 0.15 & 0.30 & 0.00 & 0.00  & 0.59  & 0.00 \\
G2 &  0.260 & 0.45 & 0.23 & 0.84 & 0.07  & 0.27 & 0.30 \\
G3 &  0.278 & 0.53 & 0.20 & 0.98 & 0.11  & 0.10  & 0.34 \\
G4 &  0.302 & 0.62 & 0.15 & 1.13 & 0.18  & 0.04 & 0.37 \\
\noalign{\vskip 2mm}   
\enddata
\end{deluxetable*}


\subsection{Effects of SFE, IMF slope, and SFH on our model predictions } \label{subsec:param}

\begin{figure*}
\includegraphics[scale=.67]{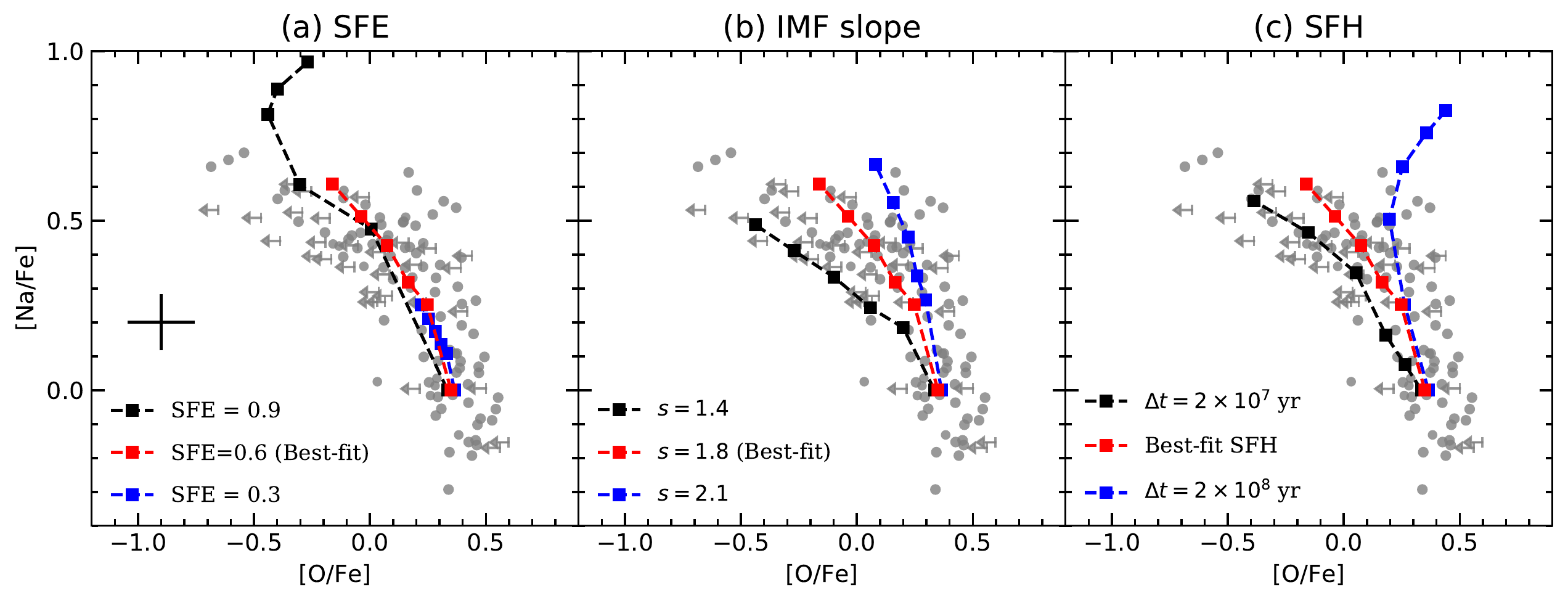}
\centering
\caption{Same as the left panel of Figure~\ref{m4m5} but to illustrate the effects of input parameters (see the text).\label{param}}
\end{figure*}

In order to illustrate the effects of variations in input parameters (SFE, IMF slope, and SFH), Figure~\ref{param} compares the observed Na-O anti-correlation of M5 with those predicted from our models constructed with different sets of parameters. For each panel, only the parameter considered is allowed to vary while other parameters are held fixed to the values of the best-fit model described above. First, the effect of SFE is shown in panel (a) where the enhancement in Na and depletion in O take place more efficiently with higher SFE. This is because, as SFE increases, a larger fraction of the gas turns into stars, and the amount of the leftover gas decreases which later mixes with the processed materials ejected from the stars. In the models with SFE of 0.9, enhancement of O, instead of depletion, is predicted at G5 and G6 because of a larger contribution from low mass AGB of G1. The effect of IMF slope is shown in panel (b). Models with top-heavier IMF predict later generations to be more depleted in O and less enhanced in Na as massive stars are responsible for O depletion while Na is mainly enriched by relatively less massive AGB. Finally, panel (c) shows the effect of time interval ($\Delta t$) between the formations of successive generations. The black and blue squares are for the cases where a fixed time interval is assumed both for the earlier and later generations. In the case with a shorter time interval (black squares; $\Delta t = 2 \times 10^{7}$ yr), only the massive stars are allowed to contribute to the chemical evolution of earlier generations, therefore, a small Na enhancement is predicted between G1 and G2 which is not enough to explain the observed gap in [Na/Fe] between the ``{Primordial}" and ``{Intermediate}" populations defined by \citeauthor{carretta09b} (\citeyear{carretta09b}; see Figure~\ref{m4m5}). The time interval of $\sim 2 \times 10^{8}$ yr between G1 and G2 is required to explain the observed gap. However, if the formation of generation continues with the same time interval (blue squares), the models fail to reproduce the O depletion at later generations because of the contribution of O enhanced ejecta from low-mass AGB of earlier generations. Therefore, in order to reproduce the observed Na-O anti-correlation, a shorter time interval ($\sim 10^{7}$ yr) is required beyond G2 as this allows the accumulation of processed materials mainly from the massive stars belonging to previous generations.
\begin{figure*}
\includegraphics[scale=.65]{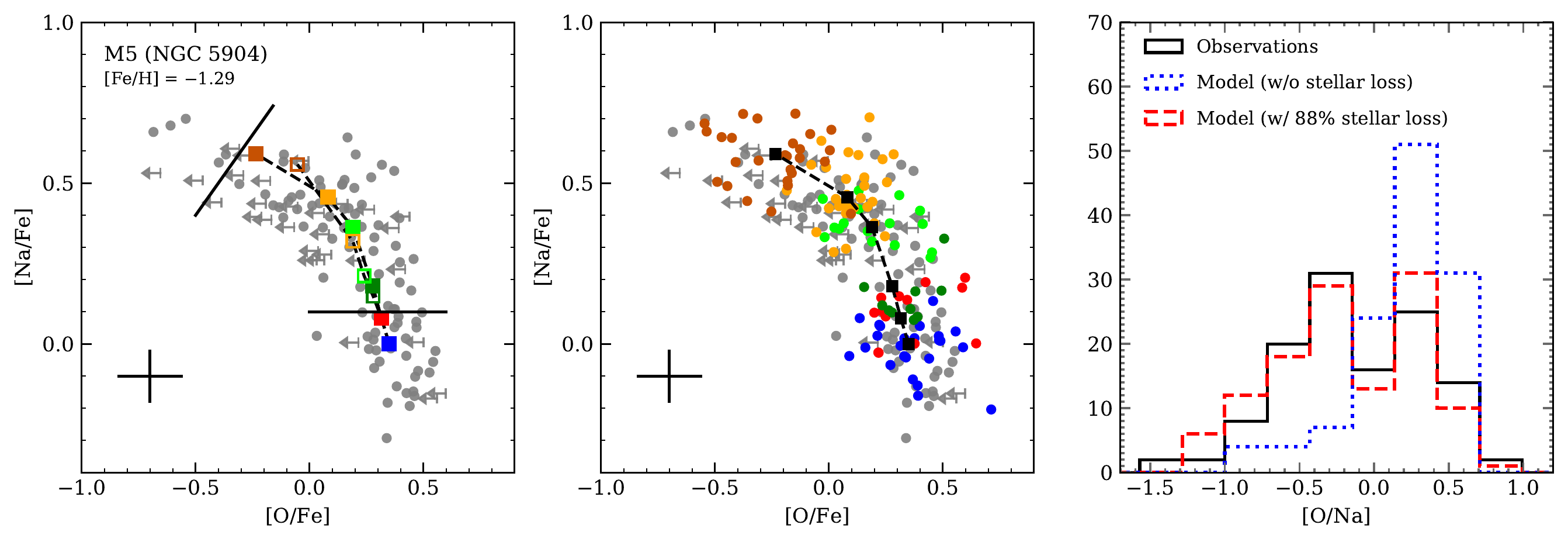}
\centering
\caption{Same as Figure~\ref{m4m5} but the predictions from the alternative model are compared. Note that this model does not match with the observed histogram unless a significant fraction ($\sim$90\%) of stars made up of earlier generations (G1, G2 \& G3) is assumed to be preferentially lost. \label{alternative}}
\end{figure*}

\begin{deluxetable*}{ccccccccc}
\setlength{\tabcolsep}{0.12in}
\tablecaption{Result of our alternative model for M5\label{tab:alternative}}
\tablehead{
\colhead{Population} & \colhead{Y} & \colhead{[Na/Fe]} & \colhead{[O/Fe]} & \colhead{[N/Fe]}  & \colhead{$\Delta$[CNO/Fe]} &\colhead{Fraction} & \colhead{Fraction} & \colhead{t(Gyr)} \\
\colhead{} &\colhead{}& \colhead{} & \colhead{} & \colhead{} & \colhead{} &\colhead{original} & \colhead{remaining} & \colhead{} 
}
\startdata
\multicolumn{3}{c}{\textbf{M~5} ([Fe/H] = $-$1.29, s = 2.1)}&&\\
\noalign{\vskip 1mm}  
G1 &  0.233 & 0.00 & 0.35 &  0.0  & 0.0  & 0.53 & 0.20 & 0.0 \\
G2 &  0.249 & 0.08 & 0.32 &  0.51 & 0.02 & 0.24 & 0.11 & 0.02 \\
G3 &  0.266 & 0.18 & 0.28 &  0.75 & 0.04 & 0.11 & 0.08 & 0.04 \\
G4 &  0.291 & 0.36 & 0.19 &  0.97 & 0.06 & 0.06 & 0.13 & 0.06 \\
G5 &  0.315 & 0.46 & 0.08 &  1.10 & 0.08 & 0.03 & 0.25 & 0.07 \\
G6 &  0.339 & 0.59 &-0.23 &  1.26 & 0.09 & 0.03 & 0.23 & 0.10 \\
\noalign{\vskip 2mm}   
\enddata
\end{deluxetable*}

\begin{figure}
\includegraphics[scale=0.7]{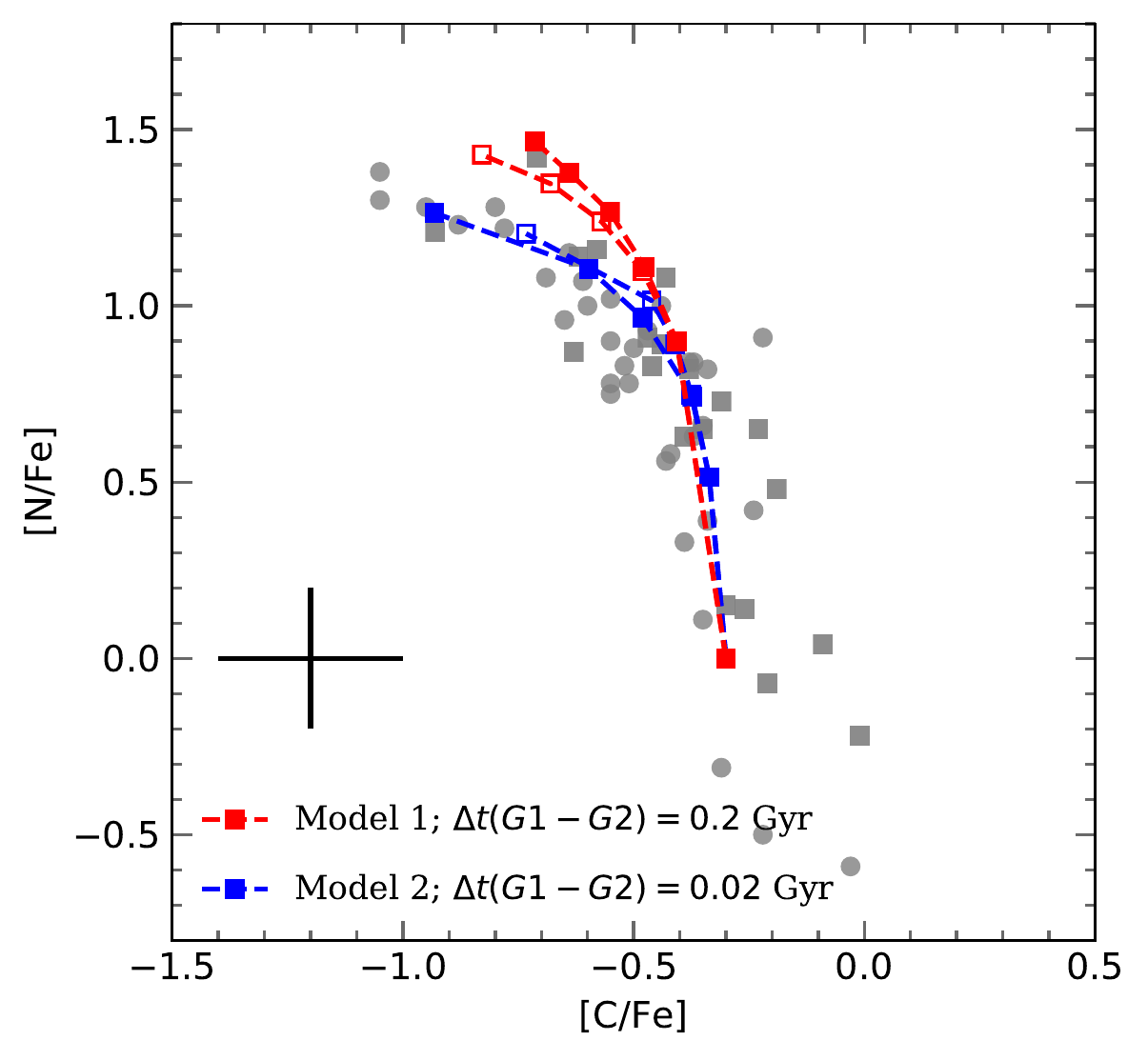}
\centering
\caption{Similar to the left panel of Figure~\ref{m4m5} but our models are compared with the observed N-C anti-correlation of M5 (data from \citealt{briley92} in gray squares; \citealt{cohen02} in gray circles). Two different models are shown where the ``{Model 1}" is from our best-fit simulation based on the Na-O anti-correlation, and the ``{Model 2}" is from the alternative model. }
\label{CN}
\end{figure}

From the fixed time interval model (with $\Delta t = 2 \times 10^{7}$ yr) in panel (c) of Figure~\ref{param}, together with the models in panel (b), one can predict that an alternative model for M5, which could also fit the observed pattern in Na-O plane, would be possible when a bottom-heavier IMF is adopted in the model. Indeed, as shown in Figure~\ref{alternative}, such a model can be found when an IMF slope of 2.1, instead of 1.8, is assumed with a quite different SFH compared to that of the best-fit model in Table~\ref{tab:m4m5} (see Table~\ref{tab:alternative}). This illustrates that the obtained SFH from the Na-O pattern is rather sensitive to the adopted IMF slope. However, in the alternative model, to match the observed [O/Na] histogram, it is required to assume that a significant fraction ($\sim$90\%) of stars  belonging to earlier generations (G1, G2 \& G3) were preferentially lost (right panel of Figure~\ref{alternative}). When normalized to the mass of the leftover stars, this fraction is an order of magnitude larger than that predicted in our best-fit models described above. For the observed pattern in N-C plane (see Figure~\ref{CN}), the alternative model (Model 2) appears to show somewhat better agreement with the observed anti-correlation compared to the best-fit model (Model 1) obtained from the Na-O anti-correlation. However, unlike Na and O abundances which are derived from atomic lines, N and C abundances are more uncertain as they are obtained by comparing the measured CN and CH molecular lines assuming a fixed O abundance \citep[e.g.,][]{cohen02, martell08, lardo12}. This, of course, is not supported from the observed depletion of O abundance in later generations (see Figure~\ref{m4m5}). Also, the observed extension of N-C anti-correlation is more prone to the deep mixing in evolved stars \citep{roediger14}. Therefore, we would obtain the best-fit parameters based on the Na-O anti-correlation including the [O/Na] histogram. Note further that, in our best-fit model, ``{Primordial}" population is made up of only G1 as would be expected from its similarity with the chemical composition of field stars, while, in the alternative model, it is composed of three subpopulations (G1, G2 \& G3) which might be considered against intuition as most field stars are expected to show G1 characteristic.  

    Our best-fit model for M5 differs  from the alternative model mainly in that the former requires a $\Delta t(G1-G2)$ which is an order of magnitude larger than that predicted in the latter model. As a consequence, the best-fit model predicts the later generations to be mildly enhanced in CNO abundances due to the contribution of CNO enhanced ejecta from low mass AGB. Such an enhancement is not predicted in the alternative model because of a longer lifetime of low mass AGB compared to the duration of star formation. Variations in CNO abundances within GCs have been reported by most of the high-resolution spectroscopic observations conducted during the last decade \citep{yong08, yong15, marino11, marino12, alves-brito12}, although more observations for ``{normal}" GCs \citep{carretta10} without heavy element spread are needed. Note, also, that a relatively large age difference along with a mild enhancement in CNO abundance are required to explain the horizontal-branch (HB) morphology and the RR Lyrae Oosterhoff period dichotomy of GCs in multiple population paradigm \citep{jang14, jang15}. Moreover, a timescale of several $10^{8}$ yr is preferred considering the gas cooling timescale, the orbital period in galactic environments which may be relevant to star formation \citep[see, e.g.,][]{conroy11}, and also the Eu enhancement observed in M5 and M15 \citep{bekki17}.

\subsection{Best-fit simulations for other globular clusters } \label{subsec:othergcs}
\begin{figure*}
\includegraphics[scale=.56]{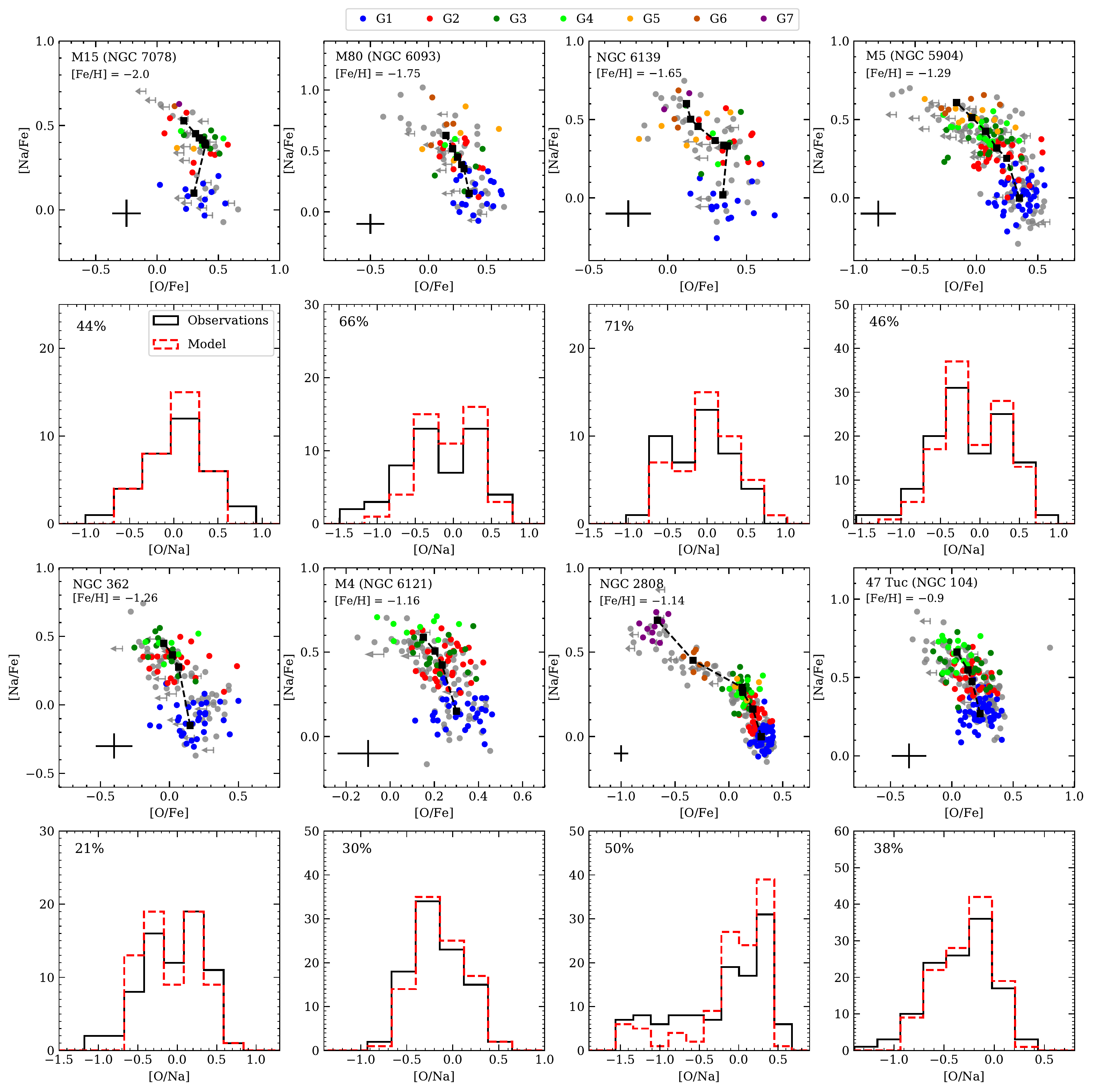}
\centering
\caption{Same as the middle and the right panels of Figure~\ref{m4m5} but including our models for other GCs \citep[data from][]{carretta09a, carretta09b, carretta13, carrettaetal15, carretta15, bragaglia15}. Eight GCs considered in this paper, including M4 and M5, are sorted in the order of increasing metallicity. For these models, full contribution of G1 ejecta is assumed to the formation of later generations. The required fraction of the lost stars is indicated in each histogram. \label{8gcs}}
\end{figure*}

Here we extend our models to other GCs to include more metal-poor and metal-rich samples ranging from M15 (NGC~7078; [Fe/H] $\approx$ $-$2.3) to 47~Tuc (NGC~104; [Fe/H] $\approx$ $-$0.7). Figure~\ref{8gcs} compares our models with the observed Na-O anti-correlations for eight GCs considered in this paper including M4 and M5. These are ``{normal}" GCs with no heavy element spread reported in the literature. The yield at specific [Fe/H] is obtained by linearly interpolating the tabulated values where the metallicity of each GC is adopted from the 2010 update of \citet{harris96} catalog with exceptions of the most metal-poor and the most metal-rich GCs (M15 and 47 Tuc). For M15, we adopt the yields of the most metal-poor stellar model ([Fe/H] = -2.0) while for 47 Tuc, we use the yields calculated at $Z=0.004$ ([Fe/H] $\approx-0.9$). For these GCs, SFHs obtained from our best-fit simulations are listed in Table~\ref{tab:othergcs}. As in M4 and M5, to better match the observed [O/Na] histograms, it is necessary to assume that some of the earlier generation stars were preferentially lost. The required fractions of the lost stars are in the range of $20-70\%$, which would imply that the original masses of the GCs to be $\sim$$1.3-3.5$ times more massive than present day values if the later generation stars remain intact. As described above, this is, however, one or two orders of magnitude smaller than those required in previous models suggested by other investigators. The discrete distributions of [O/Na] histograms in Figure~\ref{8gcs} are explained in our models by the multiple star forming episodes that are well separated with timescales ranging from $10^{7}$ to $10^{9}$ yr.

 For M15 and NGC~6139, our models suggest relatively large age differences between G1 and G2 compared to those for other GCs. Substantial enhancements in CNO abundances are therefore predicted for these GCs compared to others because of a larger contribution from low mass AGB. More observations for Na and O abundances are required, however, for these GCs in order to secure a larger sample of stars. As for the NGC~2808, an order of magnitude smaller age differences compared to other GCs are required to explain the ``{S}"-like pattern of the Na-O anti-correlation. Small age differences among G3, G4 and G5, in particular, allow the leftover gas that forms G6 to be more efficiently depleted in O by the massive stars while less massive stars are still in the main-sequence phase. It appears therefore that the variety of SFHs are needed to fit the GC data in our scenario, similar to those observed in local group dwarf galaxies although their SFHs are generally more extended \citep[see e.g.,][]{tolstoy09}.

\begin{deluxetable*}{ccccccccc}
\setlength{\tabcolsep}{0.12in}
\tablecaption{Results of our best-fit models for other GCs considered in this study\label{tab:othergcs} }
\tablehead{
\colhead{Population} & \colhead{Y} & \colhead{[Na/Fe]} & \colhead{[O/Fe]} & \colhead{[N/Fe]}  & \colhead{$\Delta$[CNO/Fe]} &\colhead{Fraction} & \colhead{Fraction} & \colhead{t(Gyr)} \\
\colhead{} &\colhead{}& \colhead{} & \colhead{} & \colhead{} & \colhead{} &\colhead{original} & \colhead{remaining} & \colhead{} 
}
\startdata
\multicolumn{3}{c}{\textbf{M15} ([Fe/H] = $-$2.0, s = 2.1)} \\
\noalign{\vskip 1mm}   
G1 &  0.230 & 0.10 & 0.30 & 0.00 & 0.00  & 0.49 & 0.34 & 0.0 \\
G2 &  0.251 & 0.39 & 0.39 & 1.10 & 0.48  & 0.27 & 0.24 & 0.90 \\
G3 &  0.261 & 0.40 & 0.39 & 1.13 & 0.51  & 0.12 & 0.21 & 0.94 \\
G4 &  0.276 & 0.41 & 0.37 & 1.19 & 0.53  & 0.06 & 0.10 & 0.95 \\
G5 &  0.291 & 0.43 & 0.34 & 1.24 & 0.55  & 0.03 & 0.05 & 0.96 \\
G6 &  0.309 & 0.45 & 0.31 & 1.30 & 0.59  & 0.01 & 0.02 & 0.97 \\
G7 &  0.357 & 0.53 & 0.22 & 1.52 & 0.74  & 0.02 & 0.04 & 1.02 \\
\noalign{\vskip 2mm}   
\hline
\multicolumn{3}{c}{\textbf{M80} ([Fe/H] = $-$1.75, s = 2.0)} \\
\noalign{\vskip 1mm}  
G1 &  0.231 & 0.15 & 0.35 & 0.00 & 0.00  & 0.48 & 0.41 & 0.00 \\
G2 &  0.256 & 0.35 & 0.31 & 1.08 & 0.14  & 0.27 & 0.24 & 0.30 \\
G3 &  0.272 & 0.39 & 0.28 & 1.24 & 0.19  & 0.12 & 0.11 & 0.33 \\
G4 &  0.293 & 0.45 & 0.25 & 1.37 & 0.25  & 0.06 & 0.05 & 0.35 \\
G5 &  0.317 & 0.52 & 0.21 & 1.49 & 0.32  & 0.04 & 0.10 & 0.37 \\
G6 &  0.346 & 0.63 & 0.15 & 1.65 & 0.44  & 0.03 & 0.09 & 0.41 \\
\noalign{\vskip 2mm}   
\hline
\multicolumn{3}{c}{\textbf{NGC~6139} ([Fe/H] = $-$1.65, s = 2.1)}\\
\noalign{\vskip 1mm}  
G1 &  0.231 & 0.02 & 0.35 & 0.00 & 0.00  & 0.50 & 0.34 & 0.00 \\
G2 &  0.253 & 0.33 & 0.38 & 1.12 & 0.27  & 0.26 & 0.26 & 0.85 \\
G3 &  0.266 & 0.34 & 0.35 & 1.20 & 0.29  & 0.11 & 0.12 & 0.89 \\
G4 &  0.287 & 0.37 & 0.30 & 1.29 & 0.31  & 0.06 & 0.06 & 0.92 \\
G5 &  0.315 & 0.46 & 0.19 & 1.45 & 0.38  & 0.04 & 0.13 & 0.97 \\
G6 &  0.330 & 0.50 & 0.15 & 1.53 & 0.42  & 0.02 & 0.06 & 0.98 \\
G7 &  0.342 & 0.60 & 0.12 & 1.62 & 0.48  & 0.01 & 0.03 & 0.99 \\
\noalign{\vskip 2mm}   
\hline
\multicolumn{3}{c}{\textbf{NGC~362} ([Fe/H] = $-$1.26, s = 2.0)} \\
\noalign{\vskip 1mm}  
G1 &  0.233 &-0.15 & 0.15 & 0.00 & 0.00  & 0.51 & 0.45 & 0.00 \\
G2 &  0.262 & 0.28 & 0.07 & 0.89 & 0.09  & 0.29 & 0.29 & 0.30 \\
G3 &  0.284 & 0.37 & 0.02 & 1.06 & 0.14  & 0.13 & 0.17 & 0.34 \\
G4 &  0.310 & 0.45 &-0.04 & 1.19 & 0.19  & 0.07 & 0.09 & 0.36 \\
\noalign{\vskip 2mm}   
\hline
\multicolumn{3}{c}{\textbf{NGC~2808} ([Fe/H] = $-$1.16, s = 1.8)}\\
\noalign{\vskip 1mm}  
G1 &  0.234 & 0.00 & 0.30 &  0.0  & 0.00  & 0.47 & 0.36 & 0.0 \\
G2 &  0.272 & 0.16 & 0.22 &  0.71 & 0.02  & 0.24 & 0.21 & 0.05 \\
G3 &  0.309 & 0.26 & 0.13 &  0.97 & 0.05  & 0.13 & 0.13 & 0.06 \\
G4 &  0.309 & 0.27 & 0.13 &  0.97 & 0.05  & 0.06 & 0.10 & 0.061 \\
G5 &  0.310 & 0.29 & 0.12 &  0.97 & 0.05  & 0.02 & 0.04 & 0.062 \\
G6 &  0.436 & 0.45 &-0.33 &  1.36 & 0.15  & 0.03 & 0.06 & 0.065 \\
G7 &  0.409 & 0.69 &-0.66 &  1.38 & 0.12  & 0.05 & 0.10 & 0.115 \\
\noalign{\vskip 2mm}   
\hline
\multicolumn{3}{c}{\textbf{47~Tuc} ([Fe/H] = $-$0.9, s = 2.1)} \\
\noalign{\vskip 1mm}  
G1 &  0.236 & 0.27 & 0.23 & 0.00 & 0.00  & 0.50 & 0.40 & 0.00 \\
G2 &  0.263 & 0.47 & 0.16 & 0.64 & 0.03  & 0.28 & 0.25 & 0.30 \\
G3 &  0.283 & 0.55 & 0.13 & 0.77 & 0.05  & 0.13 & 0.21 & 0.33 \\
G4 &  0.313 & 0.66 & 0.04 & 0.98 & 0.14  & 0.09 & 0.14 & 0.42 \\
\noalign{\vskip 2mm}   
\enddata
\end{deluxetable*}

\subsection{Model calculations with the latest yields of WMS}\label{sec:lc18}

\begin{figure*}
\includegraphics[scale=.67]{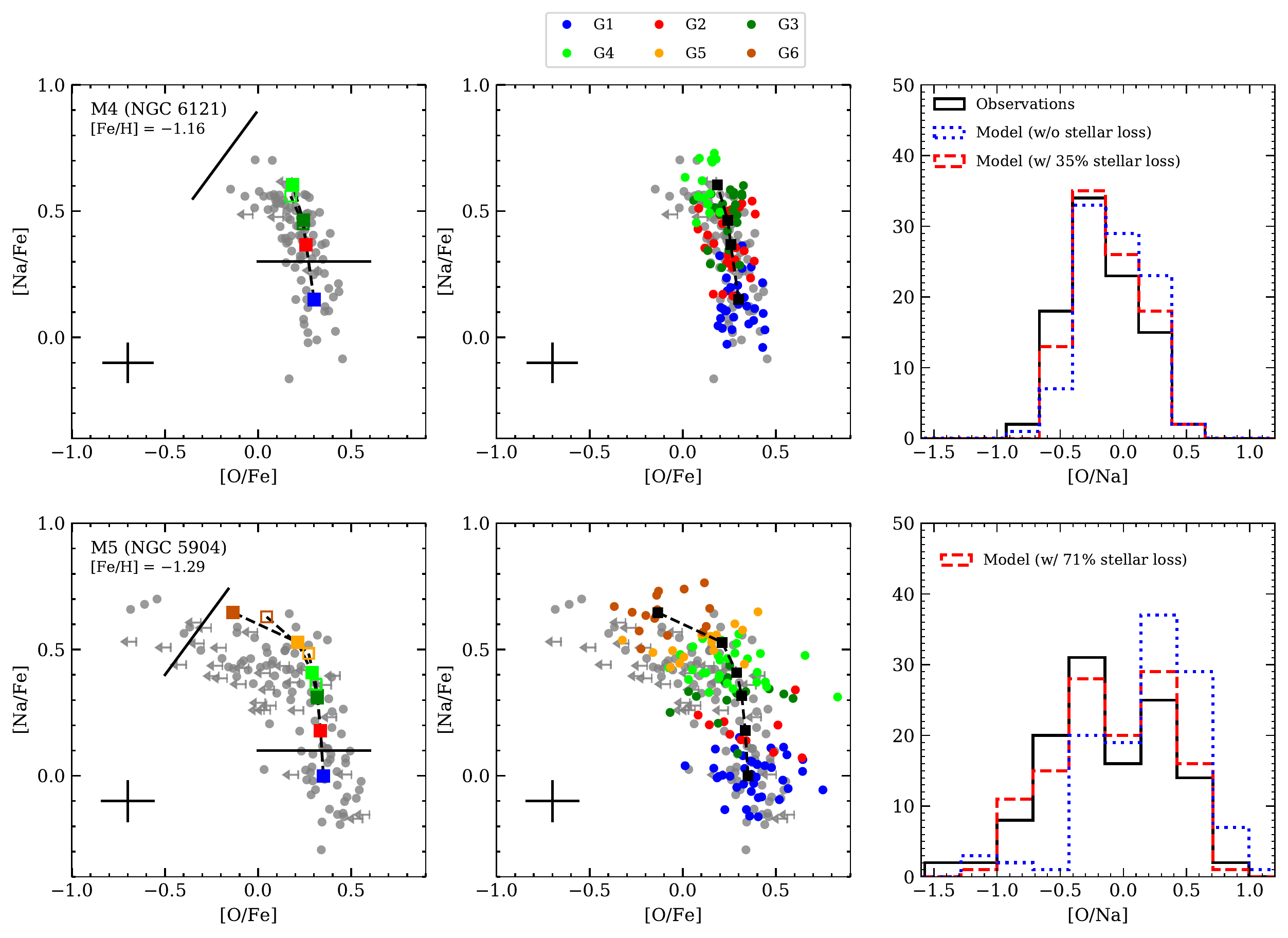}
\centering
\caption{Same as Figure~\ref{m4m5} but our models are constructed by adopting the yields of WMS without rotation from \citet{limongi18}. \label{m4m5lc18}}
\end{figure*}

\begin{figure}
\includegraphics[scale=.73]{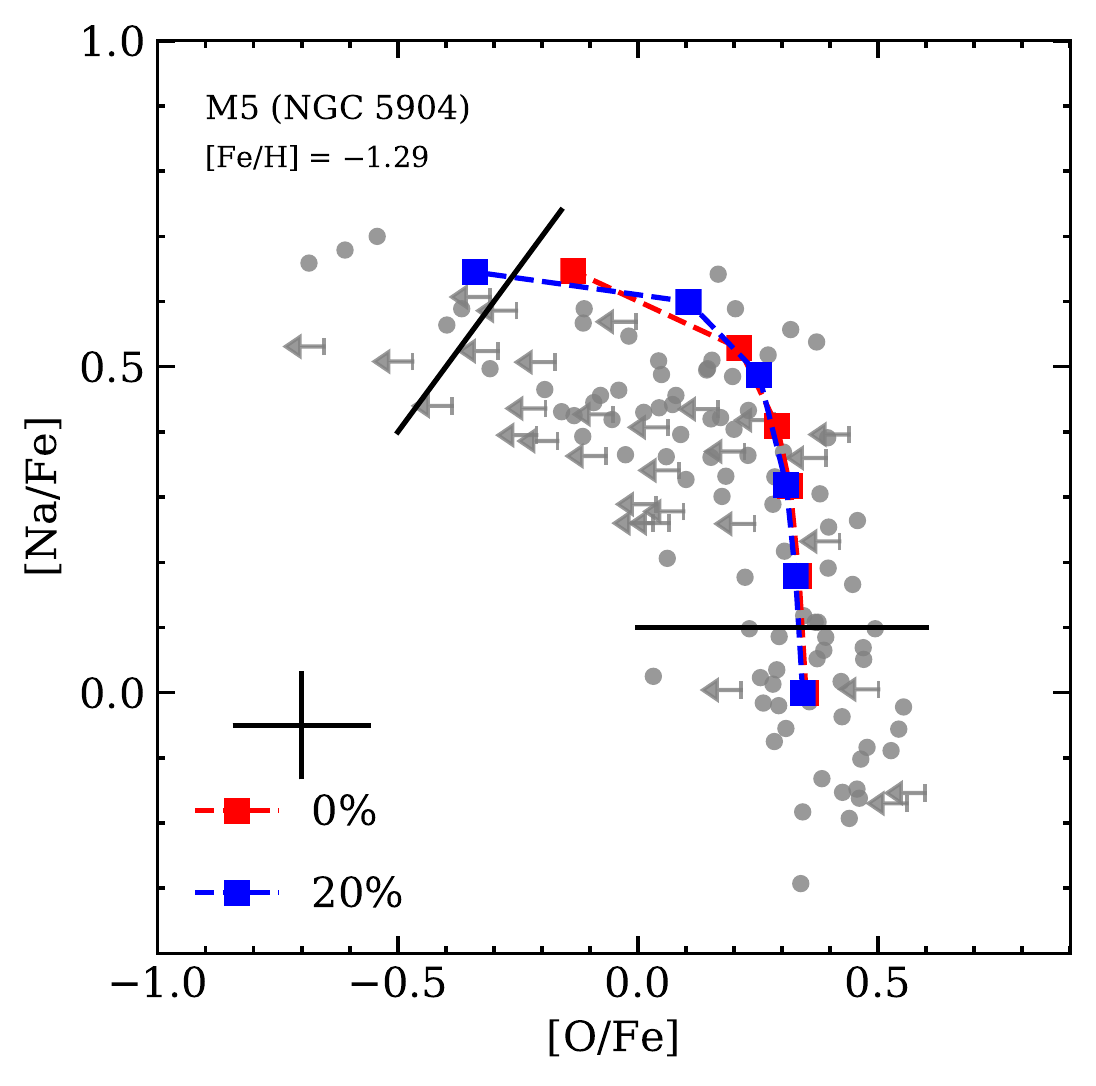}
\centering
\caption{Same as Figure~\ref{m4ent} but for M5 with the yields of WMS without rotation by \citet{limongi18}. \label{m5lc18ent}}
\end{figure}

 After this paper was submitted, we became aware of the latest calculation of massive star evolution by \citet{limongi18} where the yields of WMS are available with wide ranges of values for metallicity and rotation velocity. From this work, we have adopted metallicity dependent yields of WMS to see whether our models can still reproduce the observed Na-O anti-correlations. In Figure~\ref{m4m5lc18}, the observed Na-O anti-correlations of M4 and M5 are compared with those predicted from models constructed by adopting this new yields of massive stars without rotation. Table~\ref{tab:m4m5lc18} lists obtained SFHs from these simulations. In the case of M4, our models show a good match with the observations. The obtained SFH is similar to the one suggested above in Table~\ref{tab:m4m5} while the age difference between G1 and G2 is somewhat decreased from 0.3 Gyr to 0.2 Gyr. We find that the maximum spread of He content is decreased from  $\sim$0.07 to $\sim$0.04 which is in better agreement with that suggested by \citet{villanova12} in their measurement of He line of the blue HB stars. However, in the case of M5, the Na is overproduced by $\sim$0.1dex in the later generations and the match with observations is not as good as the models in Figure~\ref{m4m5}. The obtained SFH is also very different from that suggested in Table~\ref{tab:m4m5}. If we assume that 20\% of the ambient gas is entrained in the outflow driven by SN explosions with a shorter age difference between G5 and G6 (see Table~\ref{tab:m5lc18ent}), the match with the observed O depletion is somewhat improved (see Figure~\ref{m5lc18ent}). For NGC 2808, which displays the most extended Na-O anti-correlation, we faced a similar difficulty in reproducing the observations. It appears that  models constructed with this new set of yields do not provide good fits to the GCs  with extremely extended Na-O anti-correlations. However, we note that other GCs with moderately extended Na-O anti-correlations, like M4, are well reproduced by the new yields with SFHs similar to those suggested in Tables~\ref{tab:m4m5} and \ref{tab:othergcs}.
 
\begin{figure}
\includegraphics[scale=.6]{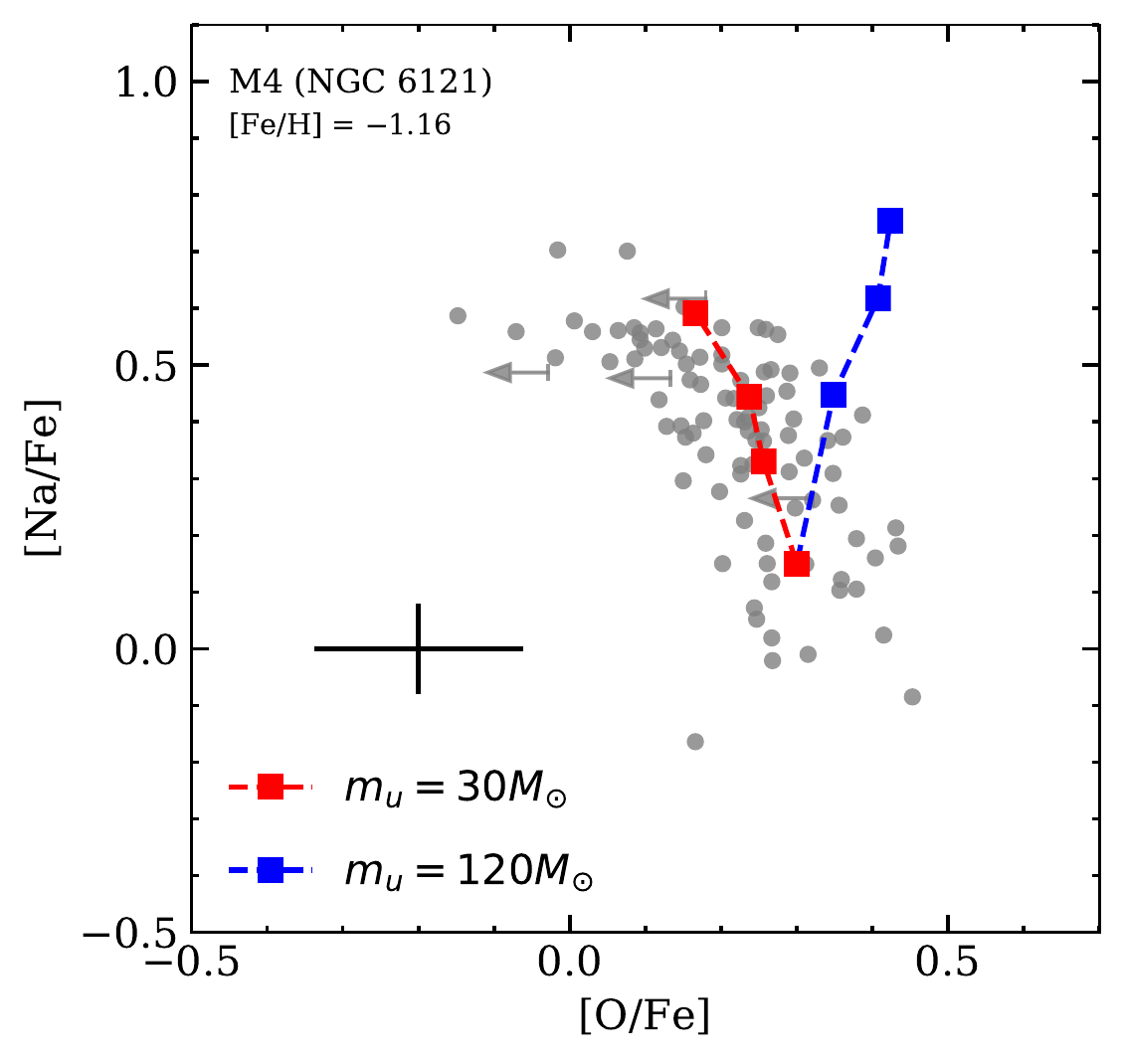}
\centering
\caption{Same as the left panel of Figure~\ref{m4m5} but the models are constructed by adopting the yields of WMS with rotation from \citet{limongi18}. In this set of yields, the O is enriched, rather than depleted, by the winds of rotating stars more massive than 30$M_{\sun}$, therefore, our models cannot match the observations (blue squares). Only when the IMF is truncated with an upper mass limit ($m_{u}$) of 30$M_{\sun}$, the observed anti-correlation is reproduced (red squares). \label{m4lc18wrot}}
\end{figure}

The effect of rotation is explored in Figure~\ref{m4lc18wrot} where the observed Na-O anti-correlation of M4 is compared with that predicted from models constructed by adopting the new yields of massive stars with rotation. The rotational velocity we employed here is not for a ``{fast-rotating}" star, but for typically observed stars with a velocity of $300~km/s$ \citep{fukuda82}. In this case, regardless of the adopted SFH, our models predict O enhancement, rather than depletion, for later generations and therefore the models cannot match the observed anti-correlation. This is because, in the models of \citet{limongi18}, winds from rotating stars more massive than $\sim$30$M_{\sun}$ are enhanced in He-burning products (C and O), even in the metal-poor regime, as they experience more efficient mixing and stronger mass-loss than non-rotating stars. Accordingly, by changing the mass range of the IMF, the observed anti-correlation might be reproduced with this new set of yields for WMS with rotation. Indeed, as shown in Figure~\ref{m4lc18wrot} (red squares) the observed correlation can be reproduced when the IMF is truncated with an upper mass limit of 30$M_{\sun}$. The obtained SFH from this model is similar to the one listed in Table~\ref{tab:m4m5lc18} (with fully populated IMF but without rotation) while the age difference between G1 and G2 is slightly decreased from 0.2 Gyr to 0.15 Gyr (see Table~\ref{tab:m4lc18wrot}). When the same truncated IMF is applied to other GCs with moderate extensions in Na-O anti-correlations, we also find SFHs similar to those suggested in Tables~\ref{tab:m4m5} and \ref{tab:othergcs}.

\begin{deluxetable*}{ccccccccc}
\setlength{\tabcolsep}{0.1in}
\tablecaption{Results of our models for M4 and M5 constructed by adopting the yields of WMS without rotation from \citet{limongi18}\label{tab:m4m5lc18}} 
\tablehead{
\colhead{Population} & \colhead{Y} & \colhead{[Na/Fe]} & \colhead{[O/Fe]} & \colhead{[N/Fe]}  & \colhead{$\Delta$[CNO/Fe]} &\colhead{Fraction} & \colhead{Fraction} & \colhead{t(Gyr)} \\
\colhead{} &\colhead{}& \colhead{} & \colhead{} & \colhead{} & \colhead{} &\colhead{original} & \colhead{remaining} & \colhead{} 
}
\startdata
\multicolumn{4}{c}{\textbf{M4} ([Fe/H] = $-$1.16, s = 2.0, SFE = 0.5)}& \\
\noalign{\vskip 1mm}  
G1 &  0.234 & 0.15 & 0.30 & 0.00 & 0.00  & 0.46 & 0.31 & 0.00 \\
G2 &  0.250 & 0.37 & 0.26 & 0.60 & 0.06  & 0.28 & 0.30 & 0.20 \\
G3 &  0.259 & 0.46 & 0.24 & 0.74 & 0.10  & 0.16 & 0.24 & 0.23 \\
G4 &  0.278 & 0.60 & 0.19 & 0.98 & 0.15  & 0.10 & 0.15 & 0.29 \\
\noalign{\vskip 2mm}   
\hline
\multicolumn{4}{c}{\textbf{M5} ([Fe/H] = $-$1.29, s = 2.1, SFE = 0.6)} \\
\noalign{\vskip 1mm}   
G1 &  0.233 & 0.00 & 0.35 & 0.00 & 0.00  & 0.55 & 0.32 & 0.00 \\
G2 &  0.241 & 0.18 & 0.34 & 0.37 & 0.05  & 0.24 & 0.08 & 0.01 \\
G3 &  0.250 & 0.32 & 0.32 & 0.59 & 0.09  & 0.07 & 0.12 & 0.04 \\
G4 &  0.260 & 0.41 & 0.29 & 0.73 & 0.11  & 0.07 & 0.23 & 0.05 \\
G5 &  0.281 & 0.53 & 0.21 & 0.93 & 0.13  & 0.03 & 0.12 & 0.06 \\
G6 &  0.317 & 0.65 &-0.13 & 1.23 & 0.13  & 0.04 & 0.13 & 0.11 \\
\noalign{\vskip 2mm}   
\enddata
\end{deluxetable*}

\begin{deluxetable*}{cccccccc}
\setlength{\tabcolsep}{0.15in}
\tablecaption{Result of our model for M5 constructed by adopting the yields of WMS without rotation from \citet{limongi18} with an assumption that 20\% of the ambient gas is entrained\label{tab:m5lc18ent}}
\tablehead{
\colhead{Population} & \colhead{Y} & \colhead{[Na/Fe]} & \colhead{[O/Fe]} & \colhead{[N/Fe]}  & \colhead{$\Delta$[CNO/Fe]} &\colhead{Fraction} & \colhead{t(Gyr)} \\
\colhead{} &\colhead{}& \colhead{} & \colhead{} & \colhead{} & \colhead{} &\colhead{original} &  \colhead{} 
}
\startdata
\multicolumn{4}{c}{\textbf{M5} ([Fe/H] = $-$1.29, s = 2.1, SFE = 0.6)} \\
\noalign{\vskip 1mm}   
G1 &  0.233 & 0.00 & 0.35 & 0.00 & 0.00  & 0.64 &  0.00 \\
G2 &  0.241 & 0.18 & 0.34 & 0.37 & 0.05  & 0.22 &  0.01 \\
G3 &  0.250 & 0.32 & 0.32 & 0.59 & 0.09  & 0.08 &  0.04 \\
G4 &  0.270 & 0.49 & 0.26 & 0.83 & 0.13  & 0.03 &  0.05 \\
G5 &  0.301 & 0.60 & 0.11 & 1.05 & 0.13  & 0.01 &  0.06 \\
G6 &  0.330 & 0.65 &-0.33 & 1.24 & 0.09  & 0.02 & 0.09 \\
\noalign{\vskip 2mm}   
\enddata
\end{deluxetable*}

\begin{deluxetable*}{cccccccc}[!]
\setlength{\tabcolsep}{0.15in}
\tablecaption{Result of our model for M4 constructed by adopting the yields of WMS with rotation from \citet{limongi18} together with a truncated IMF\label{tab:m4lc18wrot}}
\tablehead{
\colhead{Population} & \colhead{Y} & \colhead{[Na/Fe]} & \colhead{[O/Fe]} & \colhead{[N/Fe]}  & \colhead{$\Delta$[CNO/Fe]} &\colhead{Fraction} & \colhead{t(Gyr)} \\
\colhead{} &\colhead{}& \colhead{} & \colhead{} & \colhead{} & \colhead{} &\colhead{original} &  \colhead{} 
}
\startdata
\multicolumn{4}{c}{\textbf{M4} ([Fe/H] = $-$1.16, s = 2.0, SFE = 0.5)}& \\
\noalign{\vskip 1mm}  
G1 &  0.234 & 0.15 & 0.30 & 0.00 & 0.00  & 0.47 & 0.00 \\
G2 &  0.248 & 0.33 & 0.26 & 0.58 & 0.03  & 0.28 & 0.15 \\
G3 &  0.255 & 0.44 & 0.24 & 0.78 & 0.08  & 0.15 & 0.19 \\
G4 &  0.274 & 0.59 & 0.17 & 1.02 & 0.12  & 0.10 & 0.25 \\
\noalign{\vskip 2mm}   
\enddata
\end{deluxetable*}

\vskip8mm
\section{Origin of super-H\MakeLowercase{e}-rich stars in the Milky Way bulge} \label{sec:drc}

\begin{figure}
\includegraphics[scale=.5]{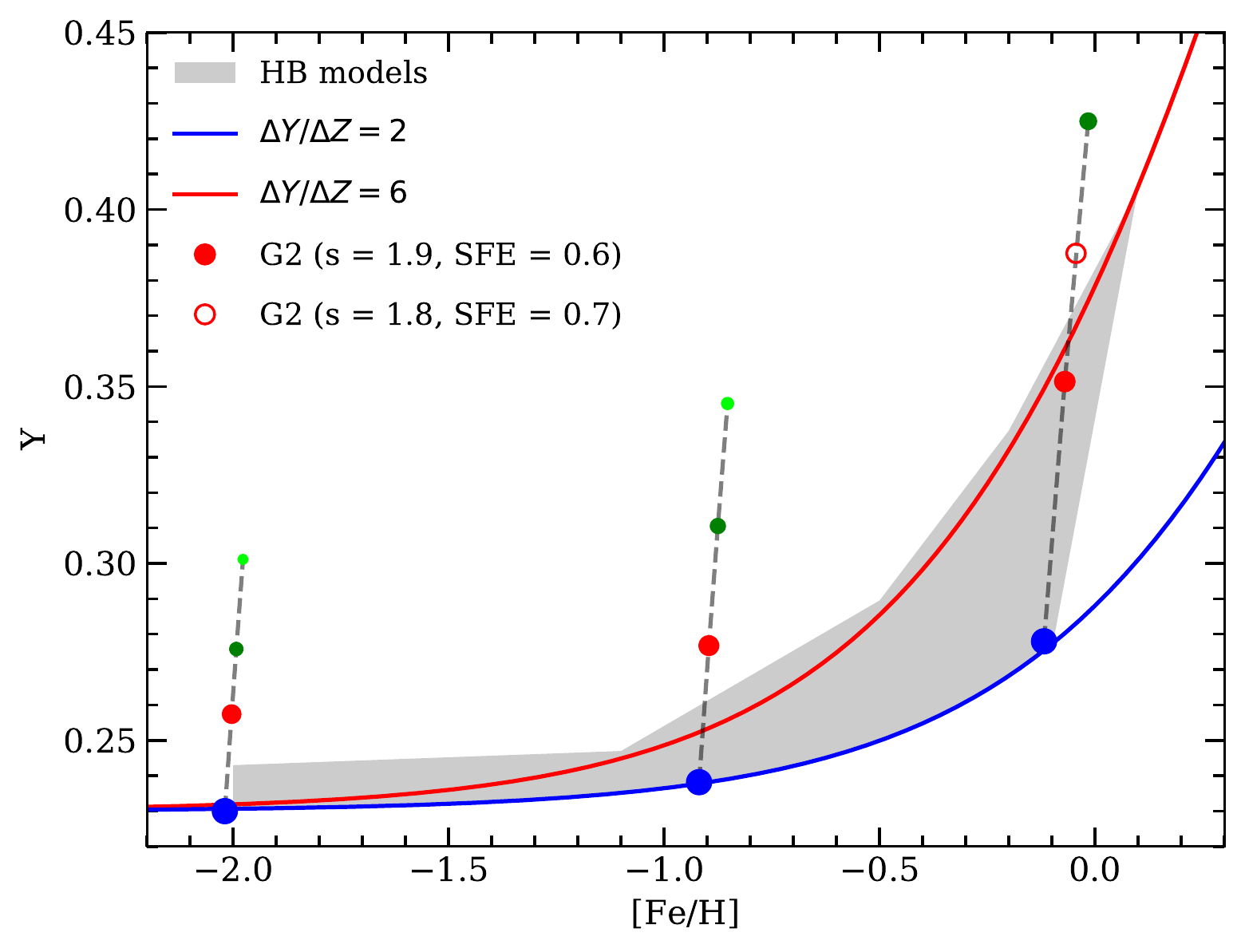}
\caption{Comparison of the predicted He contents with those estimated from synthetic HB models for G1 and G2. Our model predictions for G1, G2, G3, and G4 are shown as circles following the color scheme in Figure~\ref{m4m5}, and the size of the symbol is roughly proportional to the initial population ratio. The open circle is for the case with slightly top-heavier IMF (s = 1.8) and somewhat higher SFE (0.7). For comparison, $\Delta Y/\Delta Z = 2$ and 6 curves are also shown. Note that our model predicts a strong metallicity dependence of He enhancement between G2 and G1, which is mostly due to the metal dependent He yields of WMS (see the text). \label{Ydrc}}
\end{figure}

In this section, we have applied the same models described in previous sections to the more metal-rich subsystem in order to see whether our models can also explain the presence of the suggested super-He-rich stars in the MW bulge. In Figure~\ref{Ydrc}, the predicted He contents (Y) from our chemical evolution models are compared with those of G1 and G2 estimated from stellar evolutionary models for the HB over the full metallicity range. Specifically, we have compared the results from  synthetic HB models for the Oosterhoff dichotomy ($-$2.0 $\lesssim$  [Fe/H] $\lesssim$ $-$1.1; \citealt{jang14}, \citealt{jang15}), two populations of RR Lyrae stars in the bulge ([Fe/H] = $-$1.1; \citealt{lee16}), NGC~6388 \& NGC 6441 ([Fe/H] = $-$0.5; \citealt{caloi07}, \citealt{dantona08}), Terzan 5 ([Fe/H] = $-$0.2; \citealt{dantona10}, \citealt{joo17}), and the double RC in the MW bulge ([Fe/H] = $-$0.1; \citealt{lee15}). The adopted SFE and SFH of the models in Figure~\ref{Ydrc} are same as those of our best-fit model in Section~\ref{subsec:M4M5} for M4, which might be considered as a typical GC (see Figure~\ref{8gcs}), while a top-heavier IMF (s = 1.9) is adopted to better match the required He content for G2. Our model calculations indicate that, at given metallicity, the derived He abundance is most sensitive to the IMF slope while the effect of age difference is almost negligible. For example, the He enhancement between the subsequent generations is increased by $\Delta Y \approx 0.02$ when an IMF slope is decreased from 2.1 to 1.8 (see Tables ~\ref{tab:m4m5} and ~\ref{tab:alternative}). For the most metal-rich model at [Fe/H] $\approx$ 0.0, G4 is not shown because extremely He enhanced stars (Y $>$ 0.48) are likely to evolve directly into He-core white dwarfs instead of evolving to the RC \citep{althause17}. As for the chemical composition of G1, we assume $\Delta Y/\Delta Z = 2$, [a/Fe] = 0.3, and [N/Fe] adopted from its trend with [Fe/H] for normal stars in the inner Galaxy (\citealt{schiavon17}; see Figure~\ref{NNa} below). We also explore the case with slightly top-heavier IMF (s = 1.8) and somewhat higher value of SFE (0.7) which is shown by the open red circle in Figure~\ref{Ydrc}. This might be more relevant for the metal-rich bulge population, because, for example, \citet{tanaka18} suggested stars form more efficiently with top-heavier IMF at metal-rich regime in their theoretical investigation of massive star formation.

\begin{figure}
\includegraphics[scale=.7]{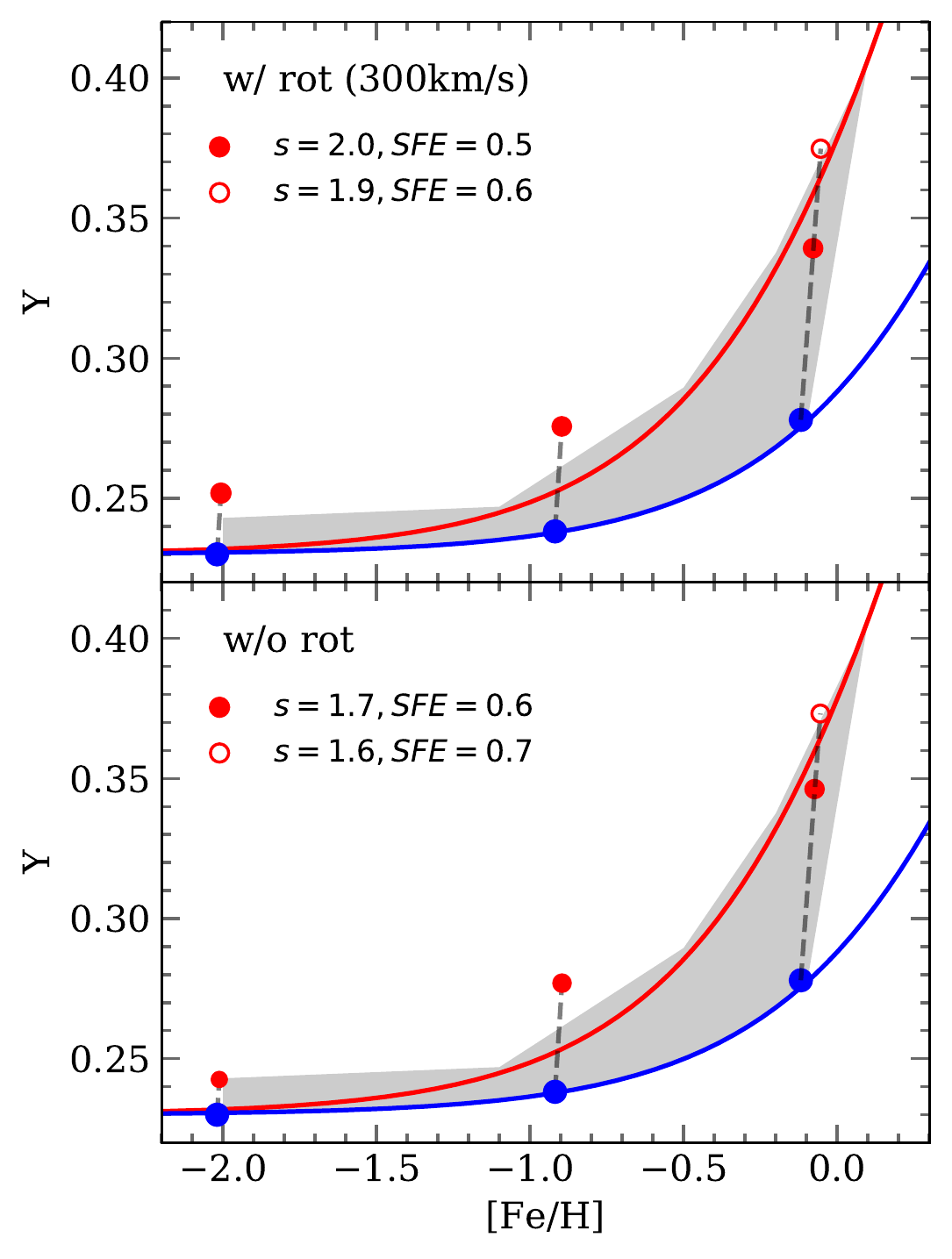}
\centering
\caption{Same as Figure~\ref{Ydrc} but our models, for G1 and G2, are constructed adopting different He yields from \citet{limongi18} for massive stars with (top panel) and without (bottom panel) rotation. While our results remain largely unchanged, a top-heavier IMF ($s = 1.7$, instead of $s=2.0$) is preferred for the models without rotation to reproduce the same He content required for the bright RC at [Fe/H] $\approx$ $-$0.1. Open symbols are for the case with slightly top-heavier IMF and somewhat higher SFE. \label{Ylc18}}
\end{figure}

\begin{figure}
\includegraphics[scale=.65]{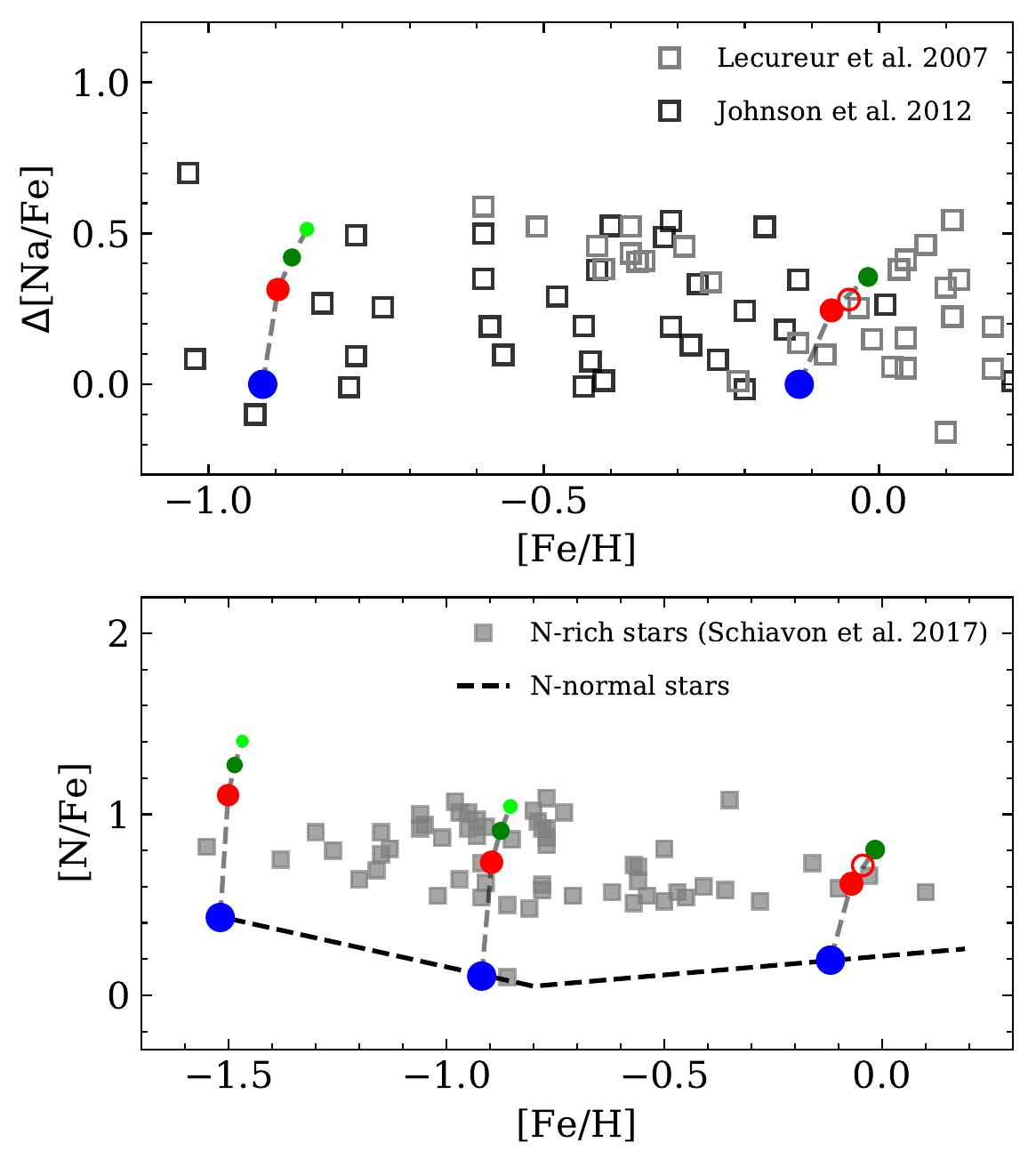}
\centering
\caption{Similar to Figure~\ref{Ydrc} but our models are compared with [N/Fe] and [Na/Fe] abundances observed in the MW bulge. In the upper panel, $\Delta$[Na/Fe] is measured with respect to the lower boundary of the observed distribution. In the lower panel, N-rich stars discovered in the inner Galaxy \citep{schiavon17} are compared with the predicted [N/Fe] abundances of G2, G3, and G4. The dashed line is the fiducial locus obtained from N-normal stars observed in the same source. \label{NNa}}
\end{figure}

\begin{deluxetable*}{ccccccccc}
\setlength{\tabcolsep}{0.12in}
\tablecaption{Results of our models at metal-rich regime ([Fe/H] = $-$0.1)\label{tab:drc}}
\tablehead{
\colhead{Population} & \colhead{Y} & \colhead{[Na/Fe]} & \colhead{[O/Fe]} & \colhead{[N/Fe]} & \colhead{$\Delta$[CNO/Fe]} & \colhead{Fraction} & \colhead{t (Gyr)}
}
\startdata
\noalign{\vskip 1mm}
G1 &  0.278 & 0.0 & 0.3 &  0.19 & 0.0 &  0.49  & 0.0 \\
G2 &  0.351 & 0.24 & 0.28 & 0.62 &  0.19 &  0.31 & 0.3 \\
G3 &  0.425 & 0.36 & 0.26 & 0.80 &  0.33 &  0.20 & 0.34 \\
\noalign{\vskip 2mm}
\hline
\noalign{\vskip 1mm}
G2 (s = 1.8, SFE = 0.7) & 0.388 & 0.28 & 0.27 &  0.72 &  0.26 &  0.31 & 0.3 \\
\noalign{\vskip 1mm}
\enddata
\end{deluxetable*}

It is clear from Figure~\ref{Ydrc} that our models predict a strong metallicity dependence of He enhancement between G1 and G2. This is mostly due to the metal-sensitive behavior of the mass-loss rate of WMS together with a larger amount of newly formed He in their yields \citep{maeder92, meynet08} as discussed in Section~\ref{subsubsec:wms}. Note that, for the same reason, WMS become major enrichment source not only for the He but also for the C, N \& O, and therefore strong depletion in C and O are not predicted at metal-rich regime in our models. This is consistent with the observations of metal-rich bulge GCs where a large spread in Na abundance is observed without a strong depletion in O \citep{munoz17, tang17}. Similarly, unlike metal-poor models, the enhancement in CNO abundance is no longer dependent on the adopted age difference because enrichments of these elements are dominated by WMS at metal-rich regime. As for the population ratio, we predict $\sim$50\% of stars in metal-rich ([Fe/H] $\approx$ $-$0.1) proto-GCs to be strongly enhanced in He (see Table~\ref{tab:drc}). Our results therefore support the suggestion by \citet{lee15} and \citet{joo17} that metal-rich proto-GCs (or subsystems similar to GCs in terms of chemical evolution) played a key role in the MW bulge formation by providing super-He-rich (G2 \& G3) and He-normal (G1) stars to the bulge field, which would be observed, respectively, as the bright and faint RCs in the bulge HR diagram. We have also constructed models adopting different He yields from the \citet{limongi18} for the massive stars with and without rotation (see Figure~\ref{Ylc18}). While our predictions of He enhancement remain largely unchanged, top-heavier IMF ($s=1.7$), instead of $s=2.0$, is preferred for the models without rotation to reproduce the same He content required for the bright RC because the He yield of WMS without rotation is smaller by $\sim$50\% compared to that with rotation.

In our models, N and Na abundances are correlated with He enhancement. However, unlike He, which is expressed in mass fraction (Y), N and Na abundances are generally expressed in relative sense with respect to the Fe abundance. Therefore, the predicted enhancements of [N/Fe] and [Na/Fe] for the later generations, at metal-rich regime, are similar to those for the metal-poor models (see Figure~\ref{NNa}). Note that these predicted enhancements are consistent with the N-rich stars discovered in the inner Galaxy \citep{schiavon17} and the observed spread in [Na/Fe] among bulge RGB stars \citep{lecureur07, johnson12}.

\begin{figure*}
\includegraphics[scale=.9]{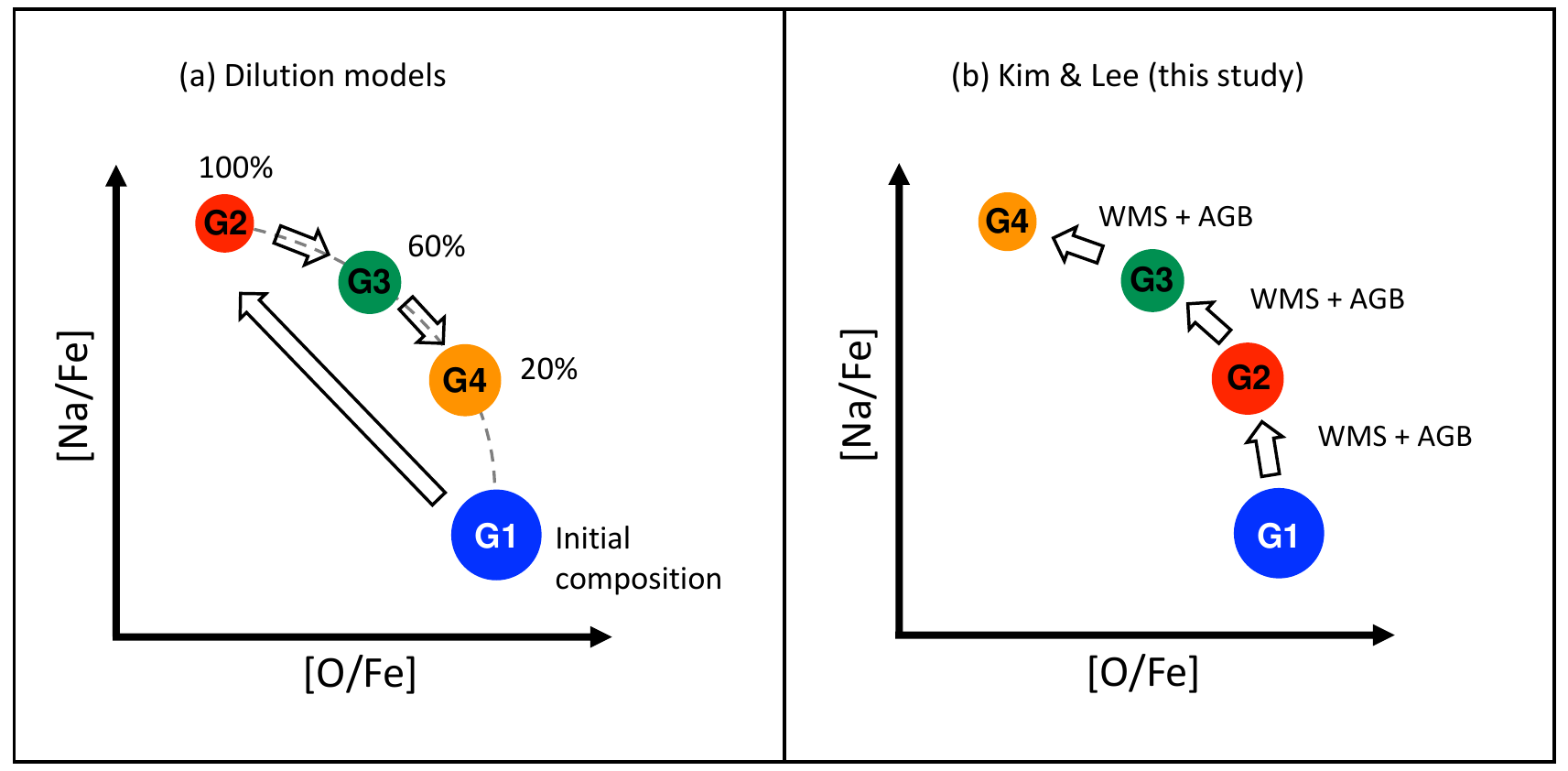}
\centering
\caption{Schematic diagram illustrating the difference between our model and previous models invoking dilution in explaining the Na-O anti-correlation. The arrows show the sequence of the formation of generations. In panel (a), the dashed line illustrates the dilution curve with the percentage roughly indicating the fraction of the processed material. Note that stars with the most extreme abundances are G2 in dilution models while those are the latest generation in our model. \label{compare}}
\end{figure*}

\section{Discussion} \label{sec:dissc}
We have presented a new chemical evolution model for proto-GCs (or low-mass subsystems similar to GCs) by adopting the two key assumptions that are different from previous approaches. One is that SN ejecta escapes the system while most of the pre-enriched ambient gas is retained as suggested by recent theoretical simulations with more realistic treatment for the proto-GC environment. The other is multiple star forming episodes with continuous enrichments by WMS and AGB from successive generations. As shown by a schematic diagram in Figure~\ref{compare}, a majority of the models suggested previously assume that stars with the most extreme abundances of Na and O are G2 which form out of the gas that is almost completely made up of the processed materials ejected by G1. The placements of the later generations on the Na-O plane are then determined by the degree of dilution of the processed materials ejected by G1 with the pristine gas. However, in our models, the enhancement (depletion) of Na (O) occurs step-by-step as the enrichments by successive generations accumulate, and therefore stars with the most extreme abundances are the latest generation. The specific pattern of Na-O anti-correlation observed in a given cluster can then be reproduced when a specific form of SFH with decreasing time interval between stellar generations is assumed.

We suspect such a unique pattern of SFH might be relevant to the degree of gas heating which delays gas cloud from collapsing and thus subsequent star formation \citep[see, e.g.,][]{rahner17}. This is because, as the generation proceeds, gas heating would be substantially weakened due to the exponentially decreasing star formation rate predicted in our models (see Table~\ref{tab:othergcs}). Furthermore, since the metal is one of the important coolants for the interstellar medium \citep{sutherland93}, gas would likely to cool down faster if the enrichment in CNO abundance continues as predicted in some of our models. Tidal perturbations caused by the orbital motions in Galactic environments may also trigger star formation with a specific history as suggested, for example, by \citet{harris04} in their investigation of the SFH of the Small Magellanic Cloud. A typical age difference between G1 and the latest generation is $\sim$0.4 Gyr in our models (see Table~\ref{tab:othergcs}). This is an order of magnitude larger than the age spreads predicted in FRMS ($\sim$$10^{7}$ yr), MIB ($\sim$$10^{7}$ yr), and massive AGB scenarios ($\sim$$10^{8}$ yr). One of the consequences of such a large age difference is the enhancement in CNO abundance which has been reported by most of the high-resolution spectroscopic observations for GCs conducted during the last decade as described above. Note again that a relatively large age difference along with some enhancement in CNO abundance are also required in stellar evolutionary models to explain the observed HB morphology (see also below). 

\begin{deluxetable*}{ccccc}
\setlength{\tabcolsep}{0.3in}
\tabletypesize{\scriptsize}
\tablecaption{Assessment of the suggested models by the observed properties of GCs\label{tab:assess}}
\tablehead{
\colhead{Models} & \colhead{GC specific} & \colhead{Variety} & \colhead{Discreteness} & \colhead{Mass budget}
}
\startdata
\noalign{\vskip 1mm}
AGB & $\bigtriangleup$ & $\bigtriangleup$ & $\bigtriangleup$ & $\bigtriangleup$\\
\noalign{\vskip 1mm}
FRMS & $\times$ & $\bigtriangleup$ & $\times$ & $\bigtriangleup$\\
\noalign{\vskip 1mm}
MIB & $\bigtriangleup$ &$\bigtriangleup$ & $\bigtriangleup$ & $\bigtriangleup$  \\
\noalign{\vskip 1mm}
SMS & $\bigtriangleup$ & $\bigtriangleup$ &  $\bigtriangleup$ & $\bigtriangleup$  \\
\noalign{\vskip 1mm}
Kim \& Lee & $\bigcirc$ & $\bigcirc$ & $\bigcirc$ & $\bigcirc$\\
\noalign{\vskip 1mm}
\enddata
\tablecomments{For the assessment of the models suggested by other investigators, we took the average from those of Table 1 of \citet{renzini15} and Figure 6 of \citet{bastian17}. Following these investigators, a cross is assigned when there is no way for the suggested model to satisfy the observational constraint; a circle is assigned if the suggested model can explain the observed property; and a triangle is assigned when the model requires ad-hoc assumptions in addition to the basic hypotheses of each model.}
\end{deluxetable*}

As summarized in Table~\ref{tab:assess}, the strongest point of our model is that the mass budget problem is mostly resolved without ad-hoc assumptions on SFE, IMF, and the preferential loss of G1. Other major observed properties of GCs are also naturally explained by our scenario. Our models are specific to GCs, as an environment similar to proto-GC is required to explain the retention of the leftover gas with the escape of SN ejecta. Discrete distributions of subpopulations observed in the HR diagram and the Na-O plane are naturally predicted as well, because our models assume multiple and discrete star forming episodes that are well separated with characteristic timescales of $10^{7}$--$10^{9}$yr. Also, by changing the details in the SFHs of GCs, the various observed patterns of Na-O anti-correlation (Figure~\ref{8gcs}) can be reproduced. In our models, GCs with very extended Na-O anti-correlation, like NGC~2808 and M5, are reproduced by invoking star forming episodes beyond G4, while GCs with modest extension in Na-O anti-correlation, such as M4, are explained by a model containing generations only up to G4. However, the previous models, as shown in the panel (a) of Figure~\ref{compare}, require some level of fine-tuning regarding the timing and the amount of pristine gas being diluted to explain the various patterns and extensions in Na-O anti-correlation as pointed out by \citet{carretta16} and \citet{bastian17}. The success of our models in reproducing the observed properties of GCs further illustrates that gas expulsion or retention is a key factor in understanding the multiple populations in GCs. Moreover, when we extend the same models to the metal-rich regime, a significant enhancement in He content between G2 and G1 is naturally predicted which is required to explain the double RC observed in the HR diagram of the MW bulge \citep{lee15, lee16}.

Most of the chemical evolution models of GCs proposed thus far mainly focused on the explanation of the anti-correlated abundance trends. However, reproducing the observed HR diagram morphology based on the multiple population paradigm is equally important. Interestingly, considering the uncertainty in the stellar yields, the parameters suggested from our chemical evolution models are already in good agreements with those obtained from synthetic HB models reproducing the HB morphology and the Oosterhoff period dichotomy of RR Lyrae stars in M15 \citep[][see their Table~1]{jang14} and in other GCs \citep{jang15}. It is unclear whether the models suggested by other investigators can also reproduce these observations as their age differences between G1 and the latest generation are much smaller ($\lesssim 10^{8}$ yr) than those predicted in our models for most GCs ($0.3-1$ Gyr). Therefore, we suggest the community to reproduce the observed abundance trends and the HR diagram morphology of GCs simultaneously as this would undoubtedly help us to better constrain the SFHs of GCs. Our forthcoming paper will address this new direction in the study of multiple stellar populations in GCs. 

\acknowledgments
We thank the referee for a number of helpful suggestions which led to several improvements in the manuscript. We also thank Dongwook Lim for his suggestion in Figure~\ref{compare}. Support for this work was provided by National Research Foundation of Korea (grants 2017R1A2B3002919 \& 2017R1A5A1070354).

\end{document}